\documentclass{article}

\usepackage[margin=1in]{geometry}
\usepackage{graphicx}
\usepackage{amsmath}
\usepackage{amssymb}
\usepackage{booktabs}
\usepackage{algorithm}
\usepackage{algpseudocode}
\usepackage{hyperref}
\usepackage{cleveref}
\usepackage{subcaption}
\usepackage{xcolor}
\usepackage[numbers,sort&compress]{natbib}
\usepackage{placeins}

\newcommand{\KE}{\mathrm{KE}}
\newcommand{\IOLweight}{e^{-\KE/kT_i}\,\sqrt{\KE}\,\sin\eta}
\newcommand{\keywords}[1]{\par\noindent\textbf{Keywords:} #1\par}

\graphicspath{{figs/}}

\title{Bayesian Active Learning of Ion Loss-Cone Boundaries in Tokamaks}

\author{
  O. E. L\'opez\thanks{email: lopezortizoe@ornl.gov}, M. Yang, M. Cianciosa, G. M. Staebler \\
  Fusion Energy Division, Oak Ridge National Laboratory \\ PO Box 2008, Oak Ridge, TN 37831, United States of America
}

\date{\today}

\begin{document}

\maketitle
\begin{abstract}
Loss-cone boundary determination in tokamaks is cast as an active learning problem in
which a Bayesian logistic regression model with radial basis function features acts as the learner and the guiding-center orbit integrator
acts as the labeling method. The approach is applied to an analytic tokamak equilibrium across a mesh of launch positions, producing a database of loss-probability models with approximate posterior uncertainty, each generated within a prescribed trajectory budget. Once generated, the database supports downstream applications, illustrated here with three examples. First, integrating the probability models against a Maxwellian distribution yields the lost-ion fraction and the energy and parallel momentum carried by lost ions, with posterior uncertainty propagated analytically to each integral. Second, a steady balance between the resulting orbit-loss torque and radial angular-momentum diffusion gives co-current edge rotation whose amplitude is set mainly by the prescribed momentum diffusivity and loss-region renewal rate. Third, a neural network surrogate trained on the database learns the in-domain mapping from launch position to loss-probability image, and transport integrals computed from its predictions closely reproduce the Bayesian reference values on held-out samples from the same mesh.
\end{abstract}


\section{Introduction}\label{sec:intro}

Ion orbit loss (IOL) is an edge mechanism with system-level consequences for tokamak performance.  When edge ions drift onto the divertor targets or first wall, they remove particles, energy and momentum, so IOL feeds into particle and edge heat-flux deposition onto plasma-facing components, intrinsic rotation, and radial electric-field formation.  The effect is consequential near the plasma boundary, where small changes in the orbit can determine whether ions remain confined or are promptly lost.  Any predictive model of IOL-driven transport depends on resolving the boundary in velocity space that separates confined from lost trajectories and tracking how that boundary changes with magnetic geometry and launch position.

Ions scattered into the velocity-space loss regions of diverted drift orbits escape across the separatrix \cite{hinton1985separatrix}.  The radial electric field modifies the low-energy loss-cone boundary \cite{chankin1993,miyamoto1996,brzozowski2019}.  Intrinsic toroidal rotation at the DIII-D H-mode edge increases with ion temperature \cite{degrassie2009intrinsic}, within the broader balance of tokamak rotation sources, transport, and sinks \cite{degrassie2009tokamak}.  Orbit loss drives a return current that sets the edge radial electric field \cite{degrassie2015thermal}, and the loss fractions grow steeply across the edge pedestal toward the separatrix \cite{stacey2011,stacey2015,wilks2016}, including losses on X-point-adjacent orbits \cite{chang2002x,ku2004}.  IOL shapes edge rotation profiles and transport fluxes in kinetic edge simulations \cite{chang2008,zhu2023}.

Given a magnetic equilibrium, a guiding-center (GC) orbit-following code can determine whether an ion launched at a specified point in phase space is lost or confined.  A bottleneck appears when that calculation must be repeated across a large number of initial conditions. Brute-force sweeps are too expensive for routine dense-mesh studies, and heuristic interpolation provides neither explicit boundary-resolution control nor posterior uncertainty estimates.

In the present work, a Bayesian logistic regression model with radial basis function (RBF) features is coupled to quadtree adaptive mesh refinement (AMR) to map loss-cone boundaries across a launch-position mesh within prescribed trajectory budgets.  Posterior uncertainty and geometric cell size guide the selection of new samples.  The resulting probability models support approximate posterior uncertainty propagation.  Locating the boundary between two outcomes of an expensive simulation is a recurring problem.  In structural engineering, for example, the boundary separates the conditions under which a structure holds from those under which it breaks \cite{bichon2008,echard2011}.  These methods adapt surrogate-based optimization \cite{jones1998} to boundary search.  They fit a Gaussian process to a continuous response and run each new simulation where the Gaussian process is least certain on which side of the boundary a point lies, so the simulations cluster along the boundary.  The loss-cone boundary in velocity space is a target of the same type.  Each GC trajectory returns one of two outcomes, so a logistic model takes the place of the Gaussian process here, and the quadtree adds explicit control of boundary resolution by subdividing only cells whose class remains uncertain.

Repeated across the launch-position mesh, these probability models form a database of Bayesian models that enables direct numerical integration against any ion distribution function, with uncertainty propagated approximately from the fitted posterior to derived quantities such as the particle and energy loss fractions.  This is the main practical advantage over a binary-label database, for which the same integration yields only a Monte Carlo average with no model-based uncertainty estimate.

The database of Bayesian models also provides supervised training targets for a neural-network surrogate that maps launch coordinates to loss-probability images.  This surrogate follows the broader use of learned emulators to accelerate expensive scientific models, including multi-fidelity and uncertainty-aware neural surrogates \cite{zhang2023multi,tatsuoka2025multifidelity,sohn2015learning,yang2025conditional,yang2025generative}.  The network is trained on the spatially resolved probability maps produced by the Bayesian workflow so that repeated in-domain queries on the same mesh family can be evaluated orders of magnitude faster than rerunning the GC-based adaptive procedure at every position.  The neural network is therefore treated as a fast interpolant over the database.

The remainder of the article is organized as follows. Sec.~\ref{sec:cerfon_equilibrium} introduces the analytic equilibrium used to define the magnetic geometry for the database. Sec.~\ref{sec:method} develops the computational method, beginning with the GC orbit model and the Bayesian AMR algorithm, then describing the single-position workflow and its boundary-resolution criteria, and closing with the mesh-wide deployment (Sec.~\ref{sec:results}) that generates the launch-position database and shows how the resulting probability maps vary across the sampled edge region.  Sec.~\ref{sec:integrals} shows how the Bayesian probability models are integrated to obtain the lost-ion fraction and normalized energy and parallel-momentum losses with propagated uncertainty, Sec.~\ref{sec:rotation} uses the inferred orbit-loss torque in a radial angular-momentum diffusion equation, and Sec.~\ref{sec:NN} describes the neural-network surrogate trained on the resulting database.  Sec.~\ref{sec:conclusions} concludes.

\section{Analytic Equilibrium}\label{sec:cerfon_equilibrium}

Ion orbit loss boundaries are sensitive to the magnetic topology near the plasma edge.  The separatrix location, X-point geometry, elongation, triangularity, and magnetic-field strength all influence whether a finite-orbit-width ion remains confined or intersects the wall.  The analytic Grad-Shafranov solutions of Cerfon and Freidberg~\cite{cerfon2010} have previously been used for GC ion orbit-loss calculations in diverted positive- and negative-triangularity tokamaks~\cite{nishimura2020}.  The present study employs the same class of solutions.  Other analytic equilibria~\cite{lopez2017high,guazzotto2021simple} could be used equally well.  Cylindrical coordinates $(R,\phi,Z)$ are employed throughout, with toroidal unit vector $\hat{\boldsymbol{\phi}}$.  In axisymmetry, the magnetic field can be written in terms of the poloidal flux $\Psi(R,Z)$ and free function $F(\Psi)$ as
\begin{equation}
  \mathbf{B} = \frac{1}{R}\nabla\Psi\times\hat{\boldsymbol{\phi}}
  + \frac{F(\Psi)}{R}\hat{\boldsymbol{\phi}}.
\end{equation}
Following Ref.~\cite{cerfon2010}, the solutions belong to the Solov'ev class, for which the pressure and toroidal-field profiles are linear in $\Psi$ and a single free parameter of that reference controls the relative pressure-gradient and toroidal-field-current contributions.  The poloidal flux is written as a particular solution plus a finite homogeneous expansion,
\begin{equation}\label{eq:cerfon_solution}
  \Psi(R,Z) = \Psi_p(R) + \sum_{n=1}^{12} c_n\Psi_n(R,Z).
\end{equation}
The particular solution $\Psi_p$ depends only on $R$ and a single free parameter.  The homogeneous solutions $\Psi_n$ are polynomials in $R$ and $Z$ with logarithmic terms in $R$.  Seven of them are even in $Z$ and five are odd in $Z$, which allows up-down asymmetric single-null configurations.  In Ref.~\cite{cerfon2010}, the coefficients $c_n$ are fixed by prescribing the flux value, tangency, and curvature of the separatrix at the outer midplane, inner midplane, upper point, and X-point.  The explicit basis functions and boundary constraints are given in that reference.

The equilibrium used to generate the Bayesian AMR database and the transport integrals (Sec.~\ref{sec:results} and Sec.~\ref{sec:integrals}) is an up-down asymmetric single-null solution of Eq.~\eqref{eq:cerfon_solution}.  The free parameter of that reference and the twelve coefficients $c_n$ are chosen to give a lower single-null diverted boundary that fits within the DIII-D wall structure, with representative tokamak-edge values of the aspect ratio, elongation, and triangularity.  The resulting boundary has minor radius $a = 0.59\,\mathrm{m}$ ($\epsilon = 0.34$), elongation $\kappa = 1.62$, triangularity $\delta = 0.35$, and a lower X-point at $(R,Z) = (1.22, -1.21)\,\mathrm{m}$.  The magnetic axis is at $(R,Z) = (1.77, 0.03)\,\mathrm{m}$, and the field normalization is set to the representative DIII-D value $B_0 = -1.87\,\mathrm{T}$.

\section{Computational Method}\label{sec:method}
\begingroup
\sloppy

Each GC orbit calculation returns a binary label in velocity space for a fixed launch position.  The ion either hits the wall within the integration window or does not.  The computational challenge is therefore to reconstruct the loss-cone boundary repeatedly across many launch points.  This section formulates that classification problem, explains why uniform sampling is inefficient, develops the Bayesian AMR algorithm used to construct probability maps with explicit boundary-resolution criteria, illustrates the workflow on a representative launch point, and generates the mesh-wide database of probability maps.

\subsection{Guiding-Center Orbit Model}

The GPU-accelerated particle pusher within the \textit{graph$\_$framework}'s code \cite{cianciosa_graphframework} integrates the relativistic GC (RGC) equations of motion for an ion of charge $Ze$ and mass $m_i$ in a prescribed magnetic field $\mathbf{B}$:
\begin{align}
  \dot{\mathbf{X}} &= \frac{1}{\hat{\mathbf{b}}\cdot\mathbf{B}^*}
    \left(Ze\,\mathbf{E}\times\hat{\mathbf{b}}
      + \frac{m_i\mu\,\hat{\mathbf{b}}\times\nabla B
        + p_\parallel\mathbf{B}^*}{m_i\gamma}\right), \label{eq:gc_pos}\\
  \dot{p}_\parallel &= \frac{\mathbf{B}^*}{\hat{\mathbf{b}}\cdot\mathbf{B}^*}
    \cdot\left(Ze\,\mathbf{E} - \frac{\mu\,\nabla B}{\gamma}\right), \label{eq:gc_mom}
\end{align}
Here $t$ is time, an overdot denotes $d/dt$, and $\mathbf{X}$ is the GC position.  The magnetic-field magnitude is $B=|\mathbf{B}|$, $\hat{\mathbf{b}} = \mathbf{B}/B$ is the unit vector along the field, $\mathbf{E}$ is the electric field (set to zero throughout this work), $e$ is the elementary charge, and $Z$ is the ion charge state.  The particle velocity and relativistic momentum are $\mathbf{v}$ and $\mathbf{p}=\gamma m_i\mathbf{v}$, with magnitudes $v=|\mathbf{v}|$ and $p=|\mathbf{p}|$.  Their components relative to the magnetic field are $v_\parallel=\mathbf{v}\cdot\hat{\mathbf{b}}$, $p_\parallel=\mathbf{p}\cdot\hat{\mathbf{b}}=\gamma m_i v_\parallel$, $v_\perp=|\mathbf{v}-v_\parallel\hat{\mathbf{b}}|$, and $p_\perp=\gamma m_i v_\perp$.  The modified magnetic field is
\begin{equation}\label{eq:gc_bstar}
  \mathbf{B}^* = Ze\,\mathbf{B} + p_\parallel\,\nabla\times\hat{\mathbf{b}}.
\end{equation}
The Lorentz factor is
\begin{equation}\label{eq:gc_gamma}
  \gamma = \sqrt{1 + \left(\frac{p_\parallel}{m_i c}\right)^2 + \frac{2\mu B}{m_i c^2}},
\end{equation}
where $c$ is the speed of light.  The magnetic moment $\mu = p_\perp^2/(2 m_i B)$ is an adiabatic invariant of the motion, and is taken as a constant in this work. Equations~\eqref{eq:gc_pos}--\eqref{eq:gc_gamma} are the exact relativistic GC form. For the ion energies considered here, however, the Lorentz factor satisfies $\gamma - 1 \lesssim 10^{-5}$ ($\KE \le 15\,\mathrm{keV}$), so the dynamics are effectively non-relativistic. In this regime the two coordinates used to label each orbit throughout this manuscript take their non-relativistic forms.  The kinetic energy is $\KE = (\gamma - 1)m_i c^2 \approx \tfrac{1}{2}m_i v^2$, and the pitch angle is $\eta = \cos^{-1}(p_\parallel/p) = \cos^{-1}(v_\parallel/v)$. Equations~\eqref{eq:gc_pos} and~\eqref{eq:gc_mom} are integrated with a fixed-step fifth-order Cash-Karp Runge-Kutta scheme~\cite{cash1990variable} throughout. The same scheme integrates the RGC equations in runaway-electron orbit calculations~\cite{beidler2020,lopez2025}.

\subsection{The Loss-Cone Classification Problem}\label{sec:loss_cone}

For a particle launched at $(R,Z)$ with kinetic energy $\KE$ and pitch angle $\eta$, the GC trajectory is advanced until either the ion intersects the wall or the integration reaches a prescribed maximum time $t_\text{max}=N_\text{steps}\Delta t$, where $N_\text{steps}$ and $\Delta t$ are the number of integration steps and integration time step, respectively.  The simulated geometry is axisymmetric, so the initial toroidal angle does not affect the orbit classification, and all simulations set $\phi(t=0)=0$ without loss of generality.  For each run, the orbit label is assigned as
\begin{equation}
  y(\KE, \eta) =
  \begin{cases}
    1 &\text{(lost: ion hits wall at $t \le t_\text{max}$)}, \\
    0 &\text{(confined: no wall hit by $t_\text{max}$)},
  \end{cases}
\end{equation}
on the domain $\Omega = [0, 15]\,\mathrm{keV} \times [0, 180]^\circ$.  The loss-cone is the set $\mathcal{L} = \{(\KE,\eta) \in \Omega : y(\KE,\eta)=1\}$ and depends on the launch position $(R,Z)$.  The goal is to locate its boundary $\partial \mathcal{L}$ at each launch position with a limited number of GC trajectories.

\begin{figure}[!htbp]
  \centering
  \begin{subfigure}[t]{0.48\linewidth}
    \includegraphics[width=\linewidth]{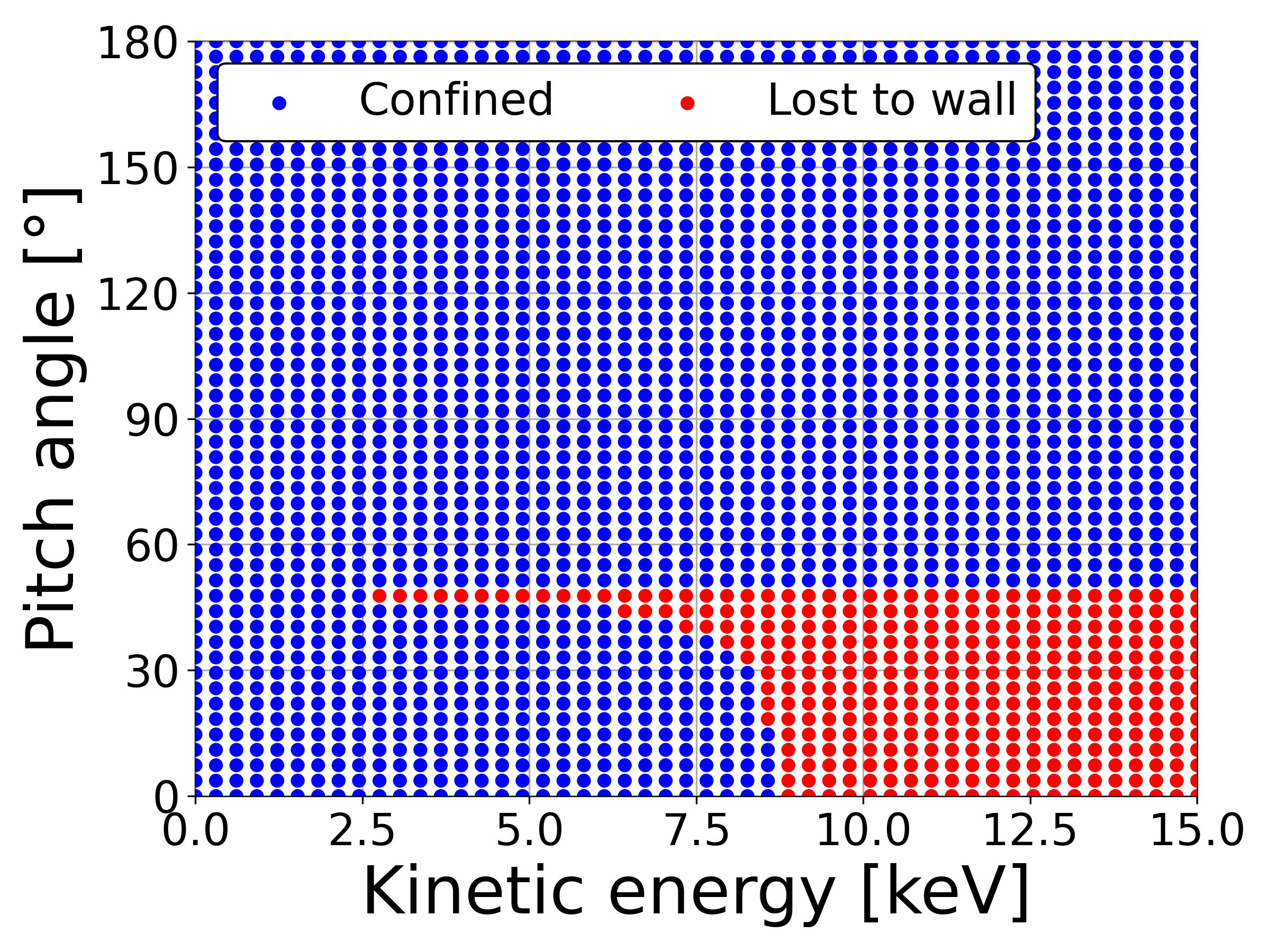}
    \caption{Each $(\KE,\eta)$ sample is labeled lost (red) or confined (blue) by direct GC orbit following.}
  \end{subfigure}
  \hfill
  \begin{subfigure}[t]{0.22\linewidth}
    \includegraphics[width=\linewidth]{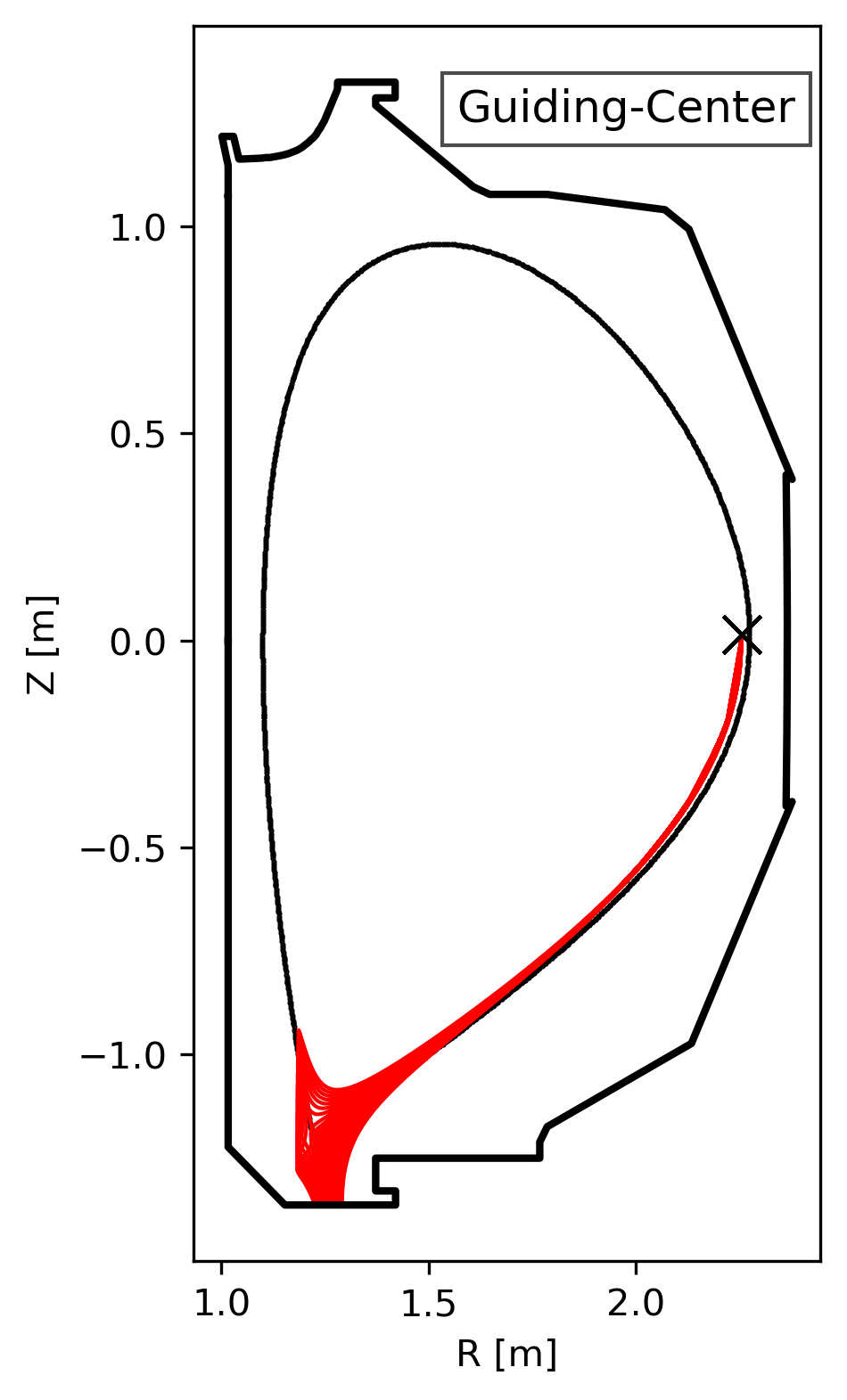}
    \caption{Lost ions reach the divertor target or first wall.}
  \end{subfigure}
  \hfill
  \begin{subfigure}[t]{0.22\linewidth}
    \includegraphics[width=\linewidth]{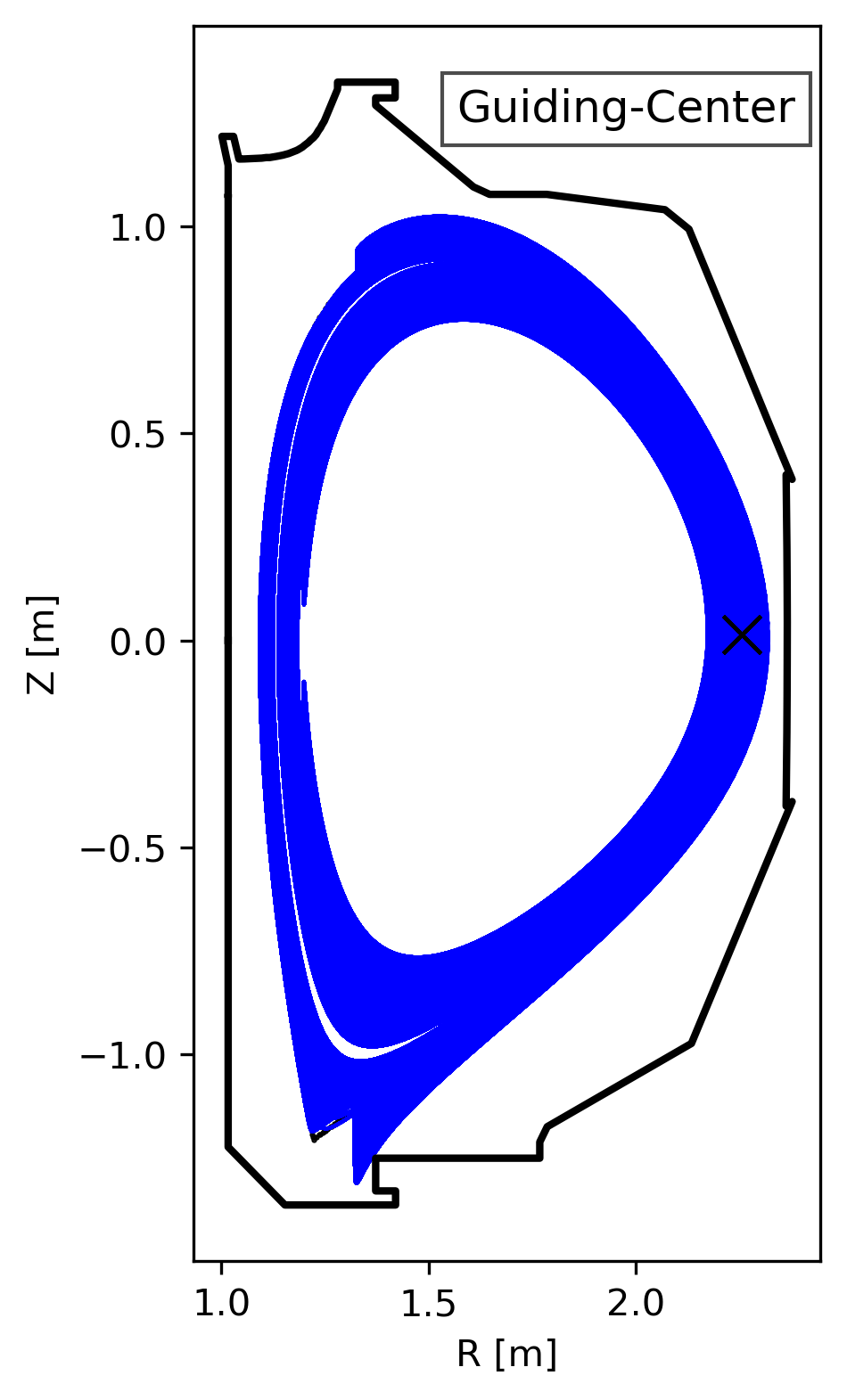}
    \caption{Confined ions: passing and banana orbits.}
  \end{subfigure}
  \caption{GC orbit classification for the equilibrium, launched from $(R, Z) = (2.26\,\mathrm{m},\, 0.01\,\mathrm{m})$.  Frame (a) shows the $50\times50$ uniform loss map over $\KE \in [0, 15]\,\mathrm{keV}$ and $\eta \in [0^\circ, 180^\circ]$.  The loss region is restricted to low pitch angles and high energies, occupying only a small fraction of the domain;
    the majority of uniformly spaced samples fall in the confined interior, where no boundary information is gained. Frames (b) and (c) show  the complete set of lost and confined trajectories in the poloidal plane.  The trajectories in this figure use $N_\text{steps}=350000$ and $\Delta t=6.55\times10^{-9}\,\mathrm{s}$, giving $t_\text{max}=2.29\,\mathrm{ms}$.}
  \label{fig:gc_classification}
\end{figure}

Fig.~\ref{fig:gc_classification} shows the inefficiency of uniform sampling.  For this representative outboard-midplane launch (denoted with an $\times$), only 336 of 2500 samples are lost, so most GC integrations land in the well-confined interior and contribute little information about the loss-cone boundary $\partial \mathcal{L}$.  The loss region is concentrated at low pitch angle and high energy. The lost ions in Fig.~\ref{fig:gc_classification}(b) and the confined ions in Fig.~\ref{fig:gc_classification}(c) illustrate the orbit families on either side of the boundary.  The target is a reconstruction of $\partial \mathcal{L}$ that concentrates new trajectories near the boundary and leaves the confined interior sparsely sampled.  Sec.~\ref{sec:bayes_amr} develops a Bayesian model that directs this sampling and also supplies a smooth loss probability for the transport integrals of Sec.~\ref{sec:integrals}.

\subsection{Bayesian Active Learning on a Quadtree Mesh}\label{sec:bayes_amr}

Each probability model is conditioned on a fixed poloidal launch position $(R,Z)$.  At that fixed launch position, the probability of loss of an ion is modeled as
\begin{equation}\label{eq:model}
  P(\text{lost} \mid \KE, \eta, w) = \varsigma(g(\KE, \eta; w)),
\end{equation}
where $\varsigma(z) = 1/(1 + e^{-z})$ is the sigmoid function, and $w = [w_0,\ldots,w_{K-1}]^\top \in \mathbb{R}^K$ is the weight vector, which holds one bias $w_0$ and $K-1$ RBF weights.  The linear predictor is
\begin{equation}\label{eq:linear_predictor}
  g(\KE, \eta; w) = w_0 + \sum_{j=1}^{K-1} w_j u_j(\KE, \eta)
  = w^\top u(\KE, \eta).
\end{equation}
Here $u(\KE,\eta) = [1,u_1(\KE,\eta),\ldots,u_{K-1}(\KE,\eta)]^\top$ is the feature vector.  Each basis function $u_j$ is a Gaussian RBF centered at $(\KE_j, \eta_j)$:
\begin{equation}\label{eq:rbf}
  u_j(\KE, \eta) = \exp\!\left(
    -\frac{(\KE - \KE_j)^2}{2\ell_{\KE}^2}
    -\frac{(\eta - \eta_j)^2}{2\ell_\eta^2}
  \right),
\end{equation}
with length scales $\ell_{\KE}$ and $\ell_\eta$ controlling the spatial extent of each feature \cite{broomhead1988}.  The centers are placed on a uniform grid of $n_{\KE}\times n_\eta$ points over $\Omega$, so $K-1 = n_{\KE}n_\eta$.  This model resolves a smooth two-dimensional classification boundary and remains cheap enough to refit after each AMR iteration.

Each sample is a point $x=(\KE,\eta)$ in $\Omega$.  Given $N$ labeled samples $\mathcal{S} = \{(x_i, y_i)\}_{i=1}^N$ with $y_i \in \{0,1\}$, the negative log-likelihood is the binary cross-entropy:
\begin{equation}\label{eq:nll}
  \mathcal{J}(w) = -\sum_{i=1}^N \left[ y_i \log p_i + (1 - y_i) \log(1 - p_i) \right],
\end{equation}
where
\begin{equation}\label{eq:pi_def}
  p_i = \varsigma(w^\top u(x_i)).
\end{equation}
A zero-mean isotropic Gaussian prior $P(w) = \mathcal{N}(0, \alpha^{-1} I_K)$ with precision $\alpha > 0$ is imposed, where $I_K$ is the $K \times K$ identity matrix.  By Bayes' theorem, the posterior over weights is $P(w \mid \mathcal{S}) \propto \exp(-\mathcal{J}(w))\,P(w)$.  The maximum a posteriori (MAP) estimate is the weight vector that maximizes this posterior, or equivalently the minimizer of the negative log-posterior:
\begin{equation}\label{eq:map}
  w_{\mathrm{MAP}} = \arg\min_w \left[ \mathcal{J}(w) + \frac{\alpha}{2} \|w\|^2 \right].
\end{equation}

The posterior $P(w \mid \mathcal{S})$ is intractable because the sigmoid likelihood is not conjugate to the Gaussian prior.  To obtain a tractable approximation with uncertainty estimates, the Laplace approximation is adopted, replacing the posterior with a Gaussian centered at $w_{\mathrm{MAP}}$ and covariance matrix $\Sigma$ \cite{mackay1992, bishop2006, seeger2004gaussian}:
\begin{equation}\label{eq:laplace}
  P(w \mid \mathcal{S}) \approx \mathcal{N}(w_{\mathrm{MAP}},\, \Sigma).
\end{equation}
Here $\Sigma$ denotes the posterior covariance matrix of the Gaussian approximation.  Under the Laplace approximation, $\Sigma$ is estimated from the Hessian $H$ of the negative log-posterior, evaluated at $w_{\mathrm{MAP}}$.  This Hessian decomposes into a data term (Fisher information) and a prior term:
\begin{equation}\label{eq:hessian}
  H = U^\top W U + \alpha I_K.
\end{equation}
Here $U \in \mathbb{R}^{N \times K}$ is the design matrix with rows $u(x_i)^\top$, and $W \in \mathbb{R}^{N \times N}$ is the diagonal matrix whose $i$th diagonal entry is $p_i(1-p_i)$, evaluated at $w_{\mathrm{MAP}}$.  In this Laplace approximation, the posterior covariance matrix is estimated as $\Sigma \approx H^{-1}$.

For a test point $x^* = (\KE^*, \eta^*)$, the linear predictor $g^* = w^\top u(x^*)$ is Gaussian under the Laplace posterior.  Its posterior mean is
\begin{equation}\label{eq:mean_score}
  \bar{g}(x^*) = w_{\mathrm{MAP}}^\top u(x^*).
\end{equation}
Its posterior variance is
\begin{equation}\label{eq:sigma}
  \sigma^2(x^*) = u(x^*)^\top \Sigma\, u(x^*),
\end{equation}
with posterior standard deviation $\sigma(x^*) = [\sigma^2(x^*)]^{1/2}$.

The refinement logic is driven by $\bar{g}$ and $\sigma$ directly, so no predictive loss probability is formed during adaptive sampling.  A credible interval for the linear predictor is constructed at any query point,
\begin{equation}\label{eq:credible}
  \mathcal{C}(x^*) = \bigl[\, \bar{g}(x^*) - \kappa_{\mathrm{acq}}\,\sigma(x^*),\;\;
                             \bar{g}(x^*) + \kappa_{\mathrm{acq}}\,\sigma(x^*) \,\bigr].
\end{equation}
The interval's position relative to zero determines whether the model classifies the query point as confidently lost, confidently confined, or uncertain.  The point is confidently lost when $\mathcal{C}(x^*)$ lies entirely above zero, confidently confined when it lies entirely below zero, and uncertain when it contains zero.

For the transport integrals reported in Sec.~\ref{sec:integrals}, the loss probability is evaluated with the MAP surrogate
\begin{equation}\label{eq:p_map}
  P_{\mathrm{MAP}}(\text{lost}\mid x^*) = \varsigma\!\bigl(\bar{g}(x^*)\bigr),
\end{equation}
and posterior uncertainty is propagated approximately from the Laplace covariance using the delta method described in Sec.~\ref{sec:integrals}.

The acquisition score follows the straddle heuristic \cite{bryan2006}, which favors points near the decision boundary ($\bar{g} = 0$) in regions of high uncertainty:
\begin{equation}\label{eq:acquisition}
  a(x) = \kappa_{\mathrm{acq}}\, \sigma(x) - |\bar{g}(x)|,
\end{equation}
where $\kappa_{\mathrm{acq}}$ sets the relative weighting between exploration (high $\sigma$) and exploitation (small $|\bar{g}|$).  The same multiplier appears in Eq.~\eqref{eq:credible}, so a single parameter fixes both the width of the credible band and the exploration weight of the acquisition score.

The Bayesian AMR run starts from a seed dataset $\mathcal{S}_0$ of $N_{\mathrm{seed}}$ trajectories launched from the corners and centers of the cells of a coarse initial mesh.  At refinement iteration $r$, the credible interval of Eq.~\eqref{eq:credible} is evaluated with the model fitted to $\mathcal{S}_{r-1}$ at the four corners and the center of each leaf cell.  A cell is assigned a confident class when all five intervals lie on the same side of zero, and is marked uncertain otherwise.  Until both classes are present among the labels, every leaf cell is marked uncertain.  An uncertain cell is subdivided when $\Delta\KE > \varepsilon_{\KE}$ or $\Delta\eta > \varepsilon_\eta$, where $\varepsilon_{\KE}$ and $\varepsilon_\eta$ are cell-width thresholds in energy and pitch angle.  After subdivision, the intervals are reevaluated on the refined mesh, and a batch of at most $k$ new trajectories is selected, integrated in one parallel GC call, and added to form $\mathcal{S}_r$.  With at most $T$ refinement iterations, a run uses at most $N_{\mathrm{seed}}+Tk$ trajectories.  The loop stops early when every uncertain cell satisfies $\Delta\KE \le \varepsilon_{\KE}$ and $\Delta\eta \le \varepsilon_\eta$, including when no uncertain cells remain.  A run that reaches iteration $T$ with uncertain cells wider than either threshold is called budget-limited.  At termination, the largest $\Delta\KE$ and $\Delta\eta$ among the remaining uncertain cells are recorded.

The batch is drawn from two candidate sources, the centers and edge midpoints of the uncertain cells that exceed either threshold and a fixed pool of random exploration points distributed uniformly over $\Omega$.  Points already evaluated are removed from both sources.  Up to a share $f_{\mathrm{explore}}$ of the $k$ slots is drawn at random from the exploration pool.  The remaining slots are filled from the cell-derived candidates by greedy farthest-point selection in $(\KE,\eta)$ coordinates rescaled to the unit square using the current candidate ranges.  Each successive candidate maximizes the minimum distance to those already chosen, and the acquisition score of Eq.~\eqref{eq:acquisition} breaks distance ties.  A quota $m$ caps the number of trajectories taken from any single cell.  The resulting batch spreads along the loss-cone boundary.  Algorithm~\ref{alg:bayes_amr} summarizes the procedure, with the run-specific hyperparameters reported in Sec.~\ref{sec:database_setup}.

\begin{algorithm}[t]
\caption{Bayesian AMR for Loss-Cone Mapping}\label{alg:bayes_amr}
\begin{algorithmic}[1]
\State \textbf{Input:} Domain $\Omega$, cell-width thresholds $\varepsilon_{\KE}, \varepsilon_\eta$, maximum refinement iterations $T$, maximum batch size $k$, exploration fraction $f_{\mathrm{explore}}$, per-cell quota $m$
\State Initialize coarse quadtree mesh; sample initial points at cell corners and centers
\State $\mathcal{S}_0 \gets$ initial labeled dataset
\State Initialize a fixed uniform random exploration pool over $\Omega$
\State $r_f \gets 0$
\For{$r = 1, \ldots, T$}
  \State Fit Bayesian logistic regression: obtain $w_{\mathrm{MAP}}$, $\Sigma = H^{-1}$
  \State Evaluate the credible interval $\mathcal{C}(x)$ at the four corners and the center of each leaf cell
  \State Mark a cell uncertain unless all five intervals lie strictly on the same side of zero; keep all cells uncertain until both classes are observed
  \If{every uncertain cell has $\Delta\KE \le \varepsilon_{\KE}$ and $\Delta\eta \le \varepsilon_\eta$}
    \State \textbf{break}
  \EndIf
  \State Mark for refinement the uncertain cells with either $\Delta\KE > \varepsilon_{\KE}$ or $\Delta\eta > \varepsilon_\eta$
  \State Subdivide marked cells (quadtree split)
  \State Reclassify leaf cells by the same rule and identify uncertain cells that exceed either threshold
  \If{no such cells remain}
    \State \textbf{break}
  \EndIf
  \State Generate candidates at those cell centers and edge midpoints; add the exploration pool and discard already evaluated points
  \State Compute acquisition scores $a(x)$ for the candidate pool
  \State Draw up to a share $f_{\mathrm{explore}}$ of the $k$ batch slots uniformly at random from the remaining exploration pool
  \State Fill the remaining slots by greedy farthest-point selection over the cell candidates, with $a(x)$ ordering equidistant candidates and at most $m$ points per cell
  \State Run GC simulations on the batch; obtain labels
  \State $\mathcal{S}_r \gets \mathcal{S}_{r-1} \cup \text{new samples}$
  \State $r_f \gets r$
\EndFor
\State Refit $(w_{\mathrm{MAP}}, \Sigma)$ to $\mathcal{S}_{r_f}$
\State Reclassify leaf cells using the final model and the same uncertainty rule
\State Measure the maximum cell widths across the remaining uncertain cells
\State \textbf{Output:} Model $(w_{\mathrm{MAP}}, \Sigma)$, MAP probability map $P_{\mathrm{MAP}}$, and maximum uncertain-cell widths $(\Delta\KE^*,\, \Delta\eta^*)$ (zero if none remain)
\end{algorithmic}
\end{algorithm}

\subsection{Single Launch-Position Demonstration}

\begin{figure}[!htbp]
  \centering
  \begin{subfigure}{0.48\linewidth}
    \includegraphics[width=\linewidth]{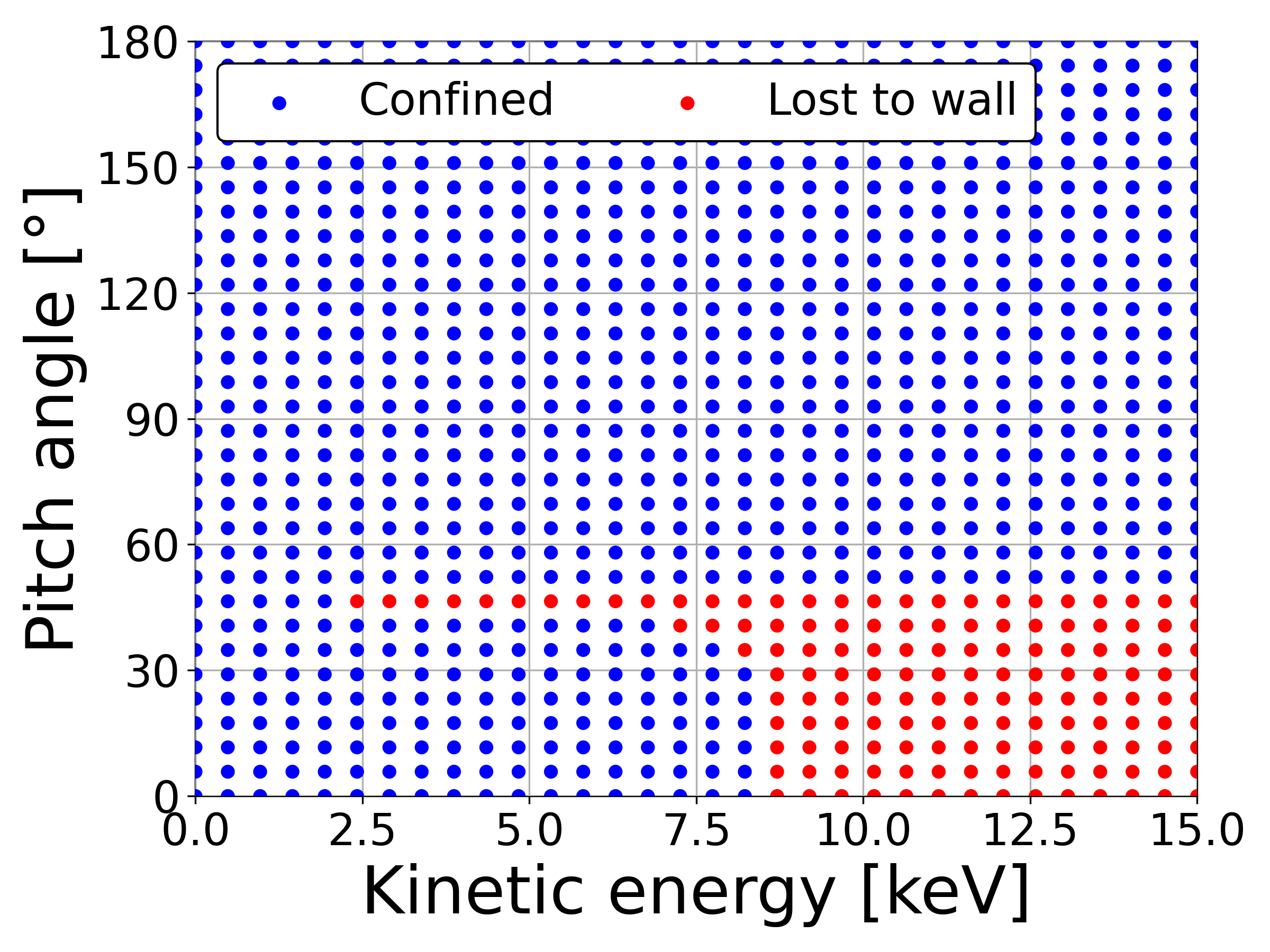}
    \caption{Uniform $32\times32$ baseline (1024~trajectories). Effort is
    distributed across the full domain.}
    \label{fig:sampling_comparison_a}
  \end{subfigure}
  \hfill
  \begin{subfigure}{0.48\linewidth}
    \includegraphics[width=\linewidth]{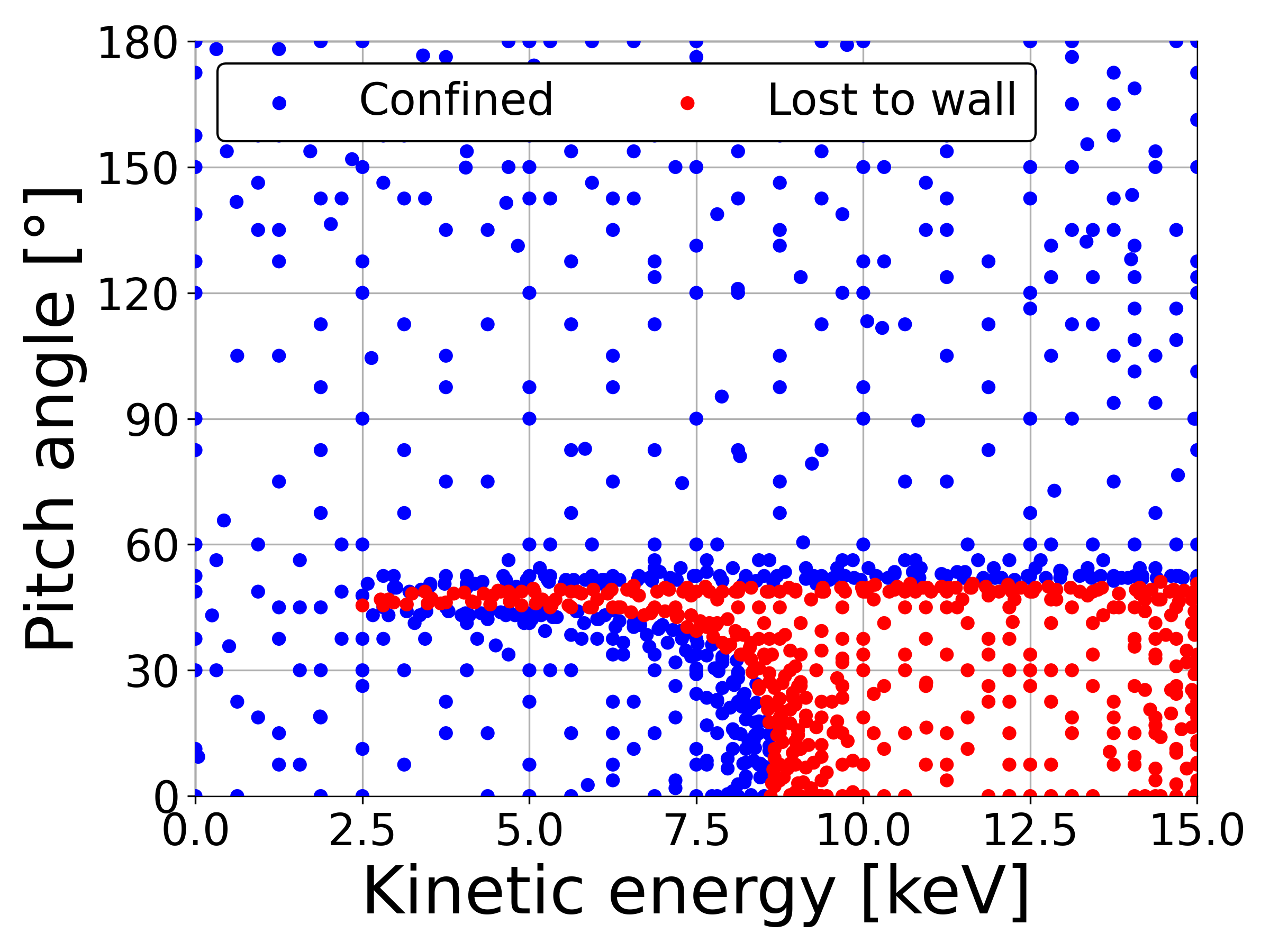}
    \caption{Bayesian AMR final sample distribution (1045~trajectories,
    7~iterations).}
    \label{fig:sampling_comparison_b}
  \end{subfigure}
  \caption{Sampling comparison at approximately equal trajectory budget.  Frame~(a) shows the
  uniform $32\times32$ baseline underlying Table~\ref{tab:comparison} metrics;
  frame~(b) shows where the Bayesian AMR method concentrates its 1045 trajectories
  after seven iterations (see Fig.~\ref{fig:iterations}).  The contrast is the central efficiency gain:
  adaptive sampling targets the loss-cone boundary.  Both calculations use $N_\text{steps}=350000$ and $\Delta t=6.55\times10^{-9}\,\mathrm{s}$, giving $t_\text{max}=2.29\,\mathrm{ms}$.}
  \label{fig:sampling_comparison}
\end{figure}

This procedure is illustrated for deuterium launches in the equilibrium from $(R, Z) = (2.26\,\mathrm{m},\, 0.01\,\mathrm{m})$.  Both methods are allocated an approximately equal trajectory budget at this launch point.  The uniform baseline evaluates a $32\times32$ mesh, while the Bayesian AMR run advances 1045 trajectories over seven iterations.  This run starts from a $6\times6$ seed mesh with 85 seed evaluations, which form iteration~1, followed by six refinement iterations of 160 trajectories each.  It uses $\kappa_{\mathrm{acq}}=2$ and an exploration pool of 48 points, with the remaining settings as in Table~\ref{tab:hyperparams}.  Under uniform sampling, effort is distributed across the full $(\KE,\eta)$ domain, and Fig.~\ref{fig:sampling_comparison}(a) shows the resulting uniform mesh.  Under Bayesian AMR, the same budget is directed toward the class transition, and Fig.~\ref{fig:sampling_comparison}(b) shows the concentration of trajectories at the loss-cone boundary after seven iterations, with the lost or confined interior left sparsely sampled.  This representative AMR run is budget-limited.  It uses all six refinement batches, and uncertain cells wider than one or both thresholds remain at termination.  The finest cells near the boundary reach $\Delta\KE=0.039\,\mathrm{keV}$ and $\Delta\eta=0.47^\circ$, a factor of 12 below the uniform spacing in each coordinate, while the coarsest cell still marked uncertain spans $0.625\,\mathrm{keV}$ and $7.5^\circ$.  The comparison therefore rests on sampling concentration, attained mesh resolution, and output type at the prescribed trajectory budget.  Table~\ref{tab:comparison} summarizes this comparison.

Fig.~\ref{fig:iterations} shows the progression of this representative active learning run from iteration~1 through iteration~7 (1045 total trajectories).  By iteration~2, the probability map has already sharpened around the class transition, the posterior uncertainty contracts toward the same band, the acquisition field concentrates on the unresolved boundary, and the set of uncertain cells becomes more localized.
At each iteration, the $P_{\mathrm{MAP}} = 0.5$ contour of Eq.~\eqref{eq:p_map}, equivalently $\bar{g} = 0$, defines the inferred loss-cone boundary, with the high-probability region concentrated at low pitch angle and higher energy and the low-energy nose retained in the final budget-limited Bayesian estimate.

\begin{table}[!t]
\centering
\caption{Comparison of sampling methods for single-position loss-cone mapping in the equilibrium.  Both methods use an approximately equal trajectory budget.}\label{tab:comparison}
\begin{tabular}{lcc}
\toprule
Metric & Uniform $32\times32$ & Bayesian AMR \\
\midrule
Total simulations            & 1024  & 1045 \\
Lost/confined labels         & 143/881 & 422/623 \\
Iterations                   & 1     & 7 \\
Sampling rule                & Uniform mesh & Uncertainty-gated quadtree AMR \\
Boundary mesh widths         & $\Delta\KE=0.48\,\mathrm{keV}$; $\Delta\eta=5.81^\circ$ & $\Delta\KE=0.039$--$0.625\,\mathrm{keV}$; $\Delta\eta=0.47^\circ$--$7.5^\circ$ \\
Model-based uncertainty      & No    & Yes \\
Output type                  & Binary & Continuous $P \in [0,1]$ \\
\bottomrule
\end{tabular}
\end{table}

\begin{figure}[!t]
  \centering
  \includegraphics[width=\linewidth]{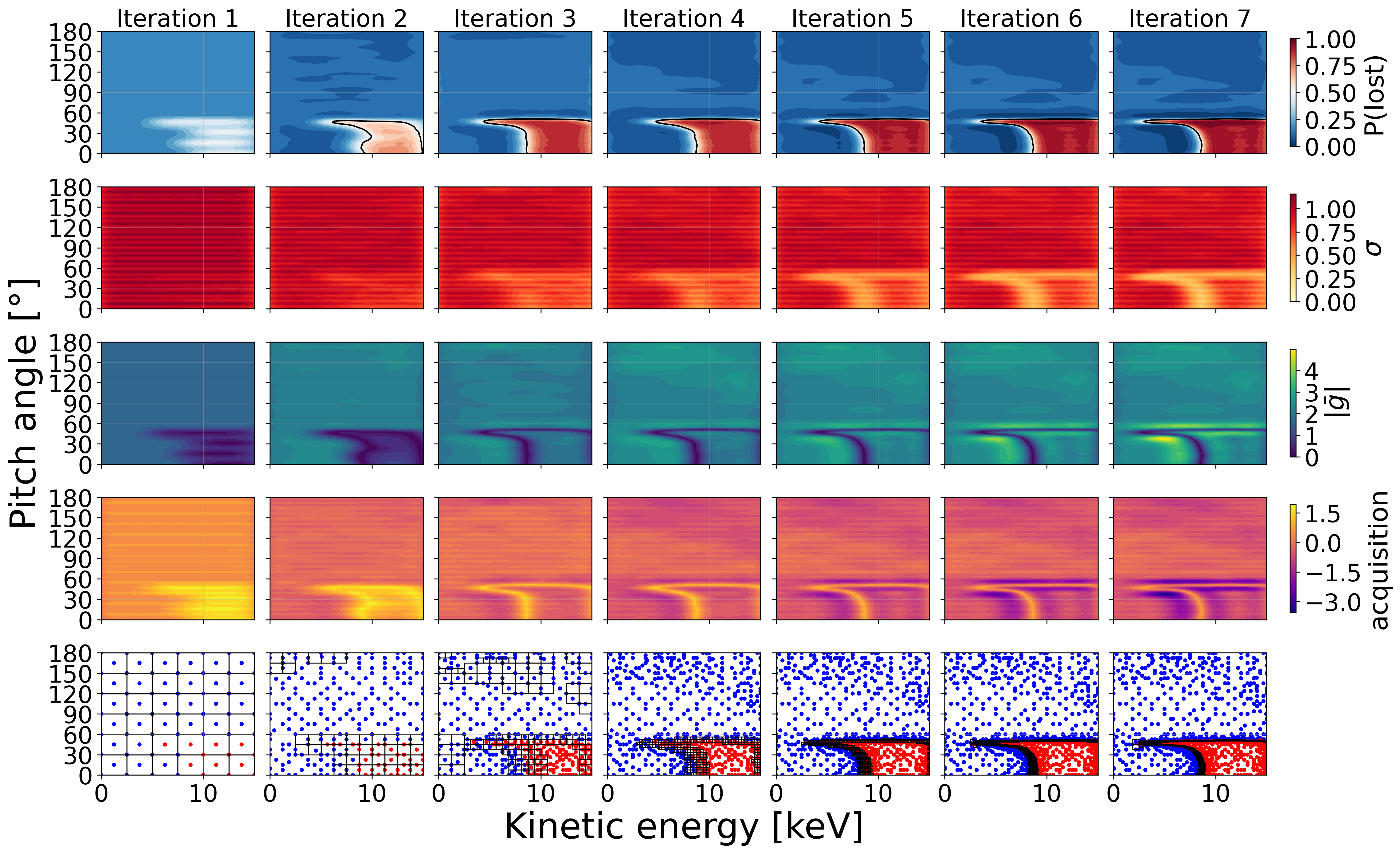}
  \caption{Active learning iteration sequence for the representative launch point
  (1045 total trajectories, 7~iterations).
  Each column corresponds to one step of the active learning loop:
  iteration~1 shows the initial Bayesian model fitted to the seed evaluations on the
  coarse corner-and-center grid; iterations~2--7 show successive updates as the acquisition function
  guides new trajectory evaluations toward the loss-cone boundary.
  The five rows give, from top to bottom, the MAP loss-probability model $P_{\mathrm{MAP}}(\text{lost})$ of Eq.~\eqref{eq:p_map},
  the posterior standard deviation $\sigma$ of the linear predictor $g$, the absolute value of its posterior mean $|\bar{g}|$
  (margin from the decision boundary), the acquisition score field, and the set of cells
  still marked uncertain, with blue (red) markers indicating the cumulative confined (lost to wall)
  trajectories evaluated through that iteration.
  The boundary-focused structure emerges after the first acquisition step and sharpens
  through subsequent iterations.  The trajectory labels use $N_\text{steps}=350000$ and $\Delta t=6.55\times10^{-9}\,\mathrm{s}$, giving $t_\text{max}=2.29\,\mathrm{ms}$.}
  \label{fig:iterations}
\end{figure}

\endgroup

\subsection{Mesh-Wide Database Generation}\label{sec:results}\label{sec:database_setup}

To compute spatially resolved IOL predictions for the present equilibrium, the single-launch calculation is repeated at launch positions distributed across the sampled edge region on a $35 \times 100 = 3500$ grid in normalized poloidal flux $\psi_N \equiv (\Psi - \Psi_\text{axis})/(\Psi_\text{sep} - \Psi_\text{axis}) \in [0.85, 0.999]$ and poloidal angle $\theta$, the standard polar angle about the magnetic axis $(R_0, Z_0)$ in the $(R,Z)$ plane, $\theta \equiv \operatorname{atan2}(Z - Z_0,\, R - R_0)$, with $\theta = 0^\circ$ at the outboard midplane and increasing counter-clockwise over $100$ values spanning $[0^\circ, 360^\circ)$.  The launch position at each $(\psi_N, \theta)$ is the first intersection of this ray with the target flux surface, so each pair maps to a unique $(R, Z)$ location.  This polar angle produces non-uniform poloidal arc-length spacing between surfaces, compressed on the inboard side.  Each launch position receives its own Bayesian AMR run, and the collection of these runs forms a spatial database of loss-probability maps.  Table~\ref{tab:hyperparams} lists the Bayesian AMR settings used to generate the database.

Each launch position is allocated $N_{\mathrm{seed}}=145$ seed trajectories and at most $T=4$ refinement batches of $k=160$ trajectories, for at most 785 trajectories.  Of the 3500 positions, 929 stop early because no cells remain uncertain.  The other 2571 complete all four refinement batches.  Of these, 2556 retain uncertain cells and 15 have none after the final model update.  The thresholds $\varepsilon_{\KE}$ and $\varepsilon_\eta$ never limit refinement within this allowance.  The mesh resolution recorded in the database is therefore set by the trajectory budget and the uncertainty classification.

\begin{table}[t]
\centering
\caption{Hyperparameters for the Bayesian AMR database generation.}\label{tab:hyperparams}
\begin{tabular}{lll}
\toprule
Parameter & Value & Description \\
\midrule
$n_{\KE}$           & 28                   & RBF centers in KE \\
$n_\eta$           & 24                   & RBF centers in pitch angle \\
$\ell_{\KE}$        & $1.2\,\mathrm{keV}$  & KE length scale \\
$\ell_\eta$        & $4.0^\circ$          & Pitch-angle length scale \\
$\alpha$             & 3.0                  & Prior precision \\
$\kappa_{\mathrm{acq}}$ & 1.0               & Acquisition and credible-band multiplier \\
Initial mesh          & $8\times8$           & Coarse AMR seed mesh \\
$N_{\mathrm{seed}}$   & 145                  & Seed trajectories on the initial mesh \\
$k$                   & 160                  & Maximum new trajectories per refinement iteration \\
$T$                   & 4                    & Maximum refinement iterations per launch position \\
Trajectory budget     & 785                  & Maximum $N_{\mathrm{seed}}+Tk$ per launch position \\
$f_{\mathrm{explore}}$ & 0.05               & Requested exploration share, subject to pool availability \\
Exploration pool      & 32                   & Initial random candidates, reused until evaluated \\
$m$                   & 12                   & Maximum batch points drawn from a single cell \\
$\varepsilon_{\KE}$  & $0.04\,\mathrm{keV}$ & KE cell-width threshold for refinement \\
$\varepsilon_\eta$ & $0.2^\circ$          & Pitch-angle cell-width threshold for refinement \\
\bottomrule
\end{tabular}
\end{table}

\begin{figure}[!htbp]
  \centering
  \includegraphics[width=\linewidth,height=0.82\textheight,keepaspectratio]{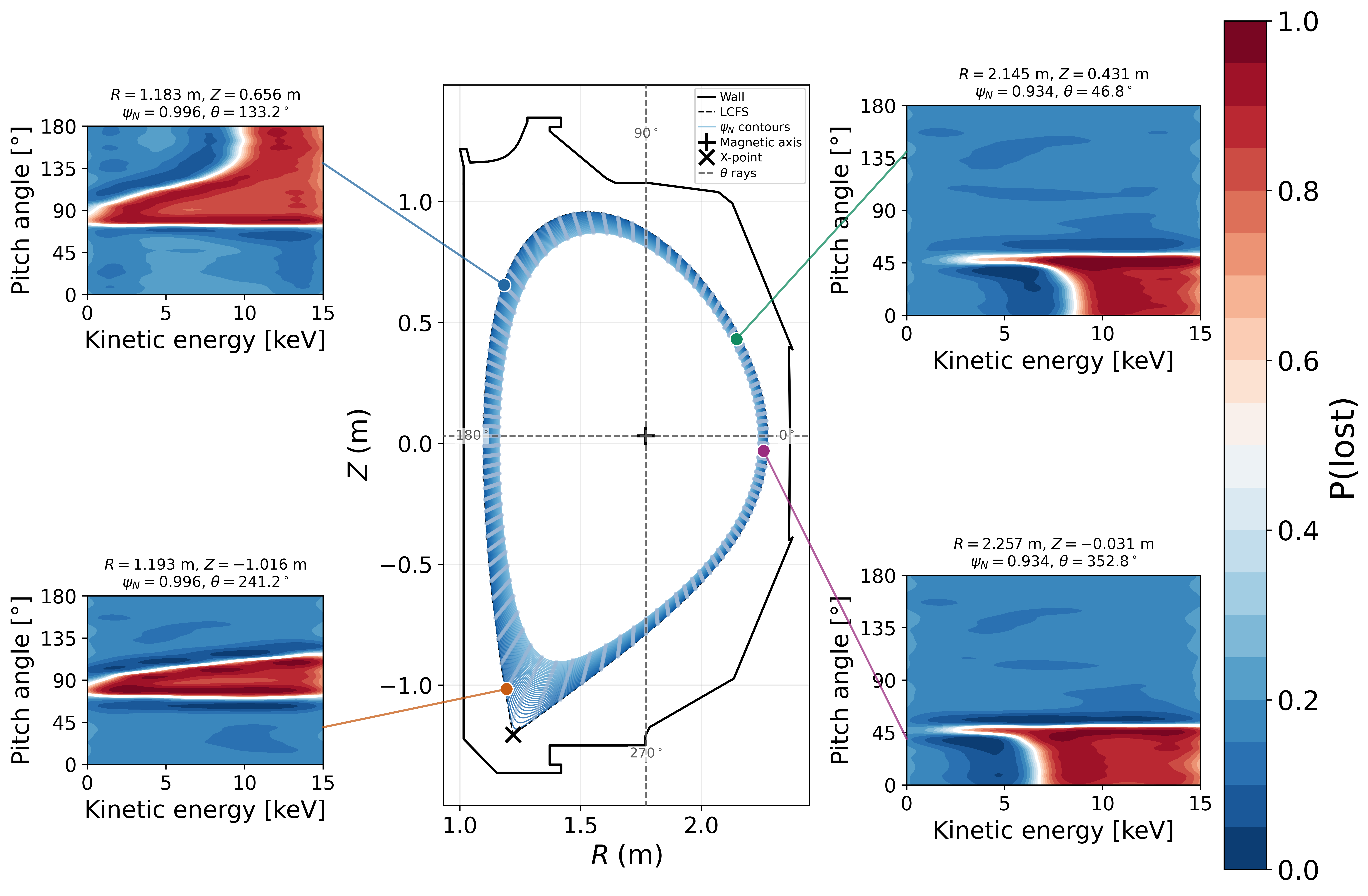}
  \caption{Representative Bayesian loss-probability maps overlaid on the launch-position mesh in $(R,Z)$ space.  The central panel shows the full $35 \times 100$ mesh-wide deployment with four highlighted edge-region launch points, while the surrounding panels show the corresponding MAP loss-probability maps in $(\KE,\eta)$ space.  The mesh-database trajectory labels use $N_\text{steps}=350000$ and $\Delta t=6.55\times10^{-9}\,\mathrm{s}$, giving $t_\text{max}=2.29\,\mathrm{ms}$.}
  \label{fig:mesh_positions}
\end{figure}

\begin{figure}[!htbp]
  \centering
  \includegraphics[width=\linewidth,height=0.85\textheight,keepaspectratio]{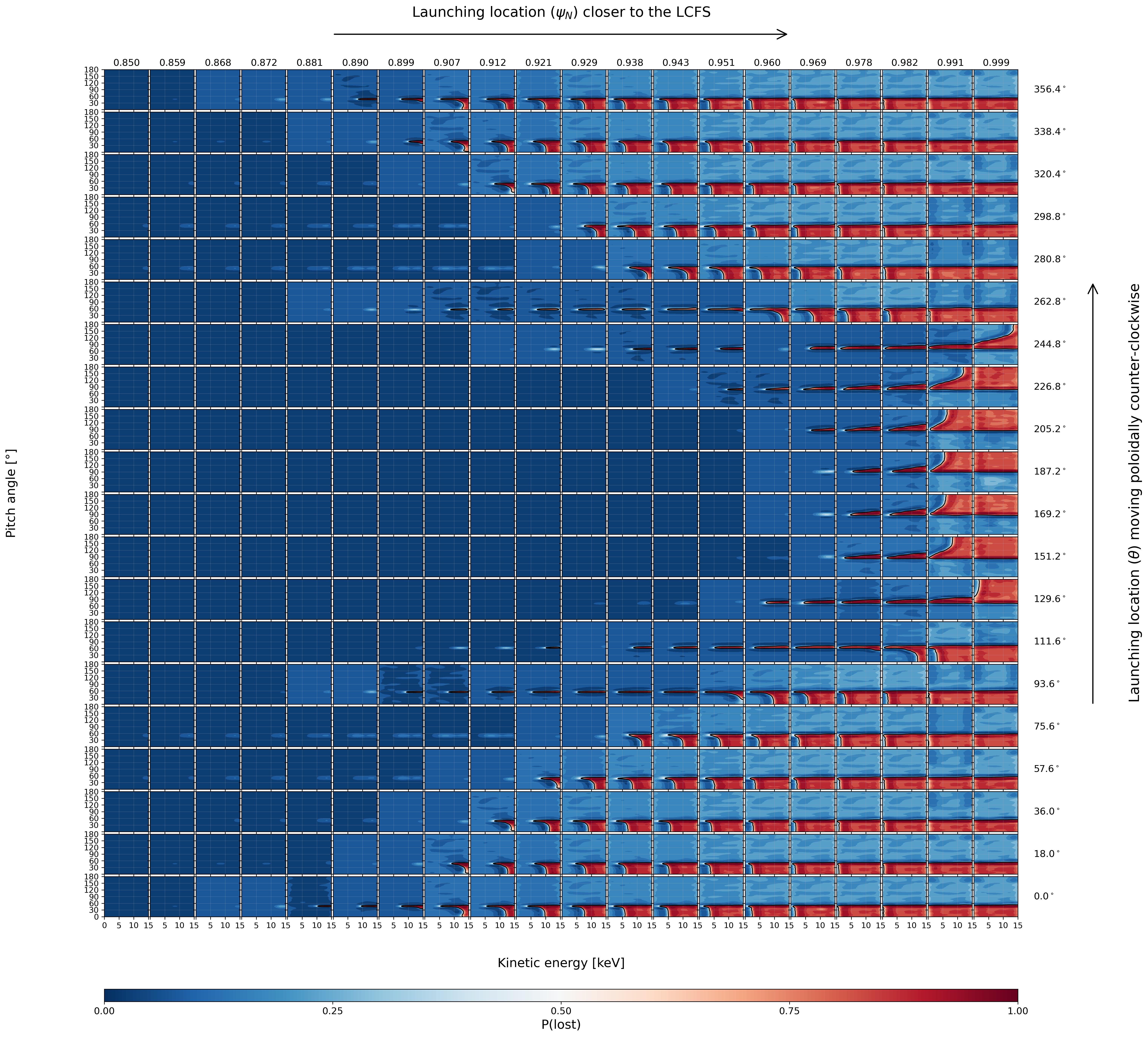}
  \caption{Representative edge-focused subset of loss-probability maps from the launch-position grid.  The $20\times20$ panel array is an evenly sampled subset of the full $35\times100$ database, spanning the sampled edge region $\psi_N \in [0.85, 0.999]$ and the full poloidal sweep; the panels illustrate how the loss-cone boundary varies with launch position in both $\psi_N$ and poloidal angle.  The full $35 \times 100$ grid constitutes the complete database generated by the present Bayesian workflow and supplies the training, validation, and test data for the neural network surrogate of Sec.~\ref{sec:NN}.  The trajectory labels use $N_\text{steps}=350000$ and $\Delta t=6.55\times10^{-9}\,\mathrm{s}$, giving $t_\text{max}=2.29\,\mathrm{ms}$.}
  \label{fig:no_efield_grid}
\end{figure}

Fig.~\ref{fig:mesh_positions} shows the geometric distribution of the launch database in the sampled edge region together with four representative probability maps taken from highlighted edge-region launch points.  Fig.~\ref{fig:no_efield_grid} then shows a broader edge-focused subset of the database, an evenly sampled $20\times20$ array spanning the sampled edge region $\psi_N \in [0.85, 0.999]$ and the full poloidal sweep.  This mesh resolves the edge region while keeping a regular grid in $(\psi_N,\theta)$ for downstream aggregation and surrogate training, and the probability maps in both figures show how the inferred loss-cone boundary changes with launch position.

\section{IOL Transport Integrals}\label{sec:integrals}

The Bayesian surrogate produced at each launch position provides a continuous representation of the loss-cone that can be integrated directly against any ion distribution function.  The integrals below take the loss probability from the MAP surrogate of Eq.~\eqref{eq:p_map} evaluated at $x^*=(\KE,\eta)$, so $P(\text{lost}\mid\KE,\eta)$ denotes $P_{\mathrm{MAP}}$, and posterior uncertainty is propagated approximately from the Laplace covariance by the delta method.  Three dimensionless transport moments are reported, each averaged over the full Maxwellian and therefore expressed per background ion.  They are the lost-ion fraction $\widetilde{\mathcal{N}}_\text{loss}$, the energy carried by lost ions $\widetilde{\mathcal{E}}_\text{loss}$ normalized by $\tfrac{3}{2}kT_i$, and the parallel momentum carried by lost ions $\widetilde{\mathcal{M}}_\text{loss}$ normalized by $p_\text{th}=\sqrt{2\,m_i\,kT_i}$.  Here $kT_i$ is the ion temperature in energy units, $m_i$ is the ion mass, and $p_\text{th}$ is the corresponding thermal momentum scale.  The sign of $\widetilde{\mathcal{M}}_\text{loss}$ tracks the co/counter-current asymmetry of the lost population, which is relevant to intrinsic rotation and return-current balance in the edge \cite{degrassie2015thermal,chankin1993,piper2019}.

The moments in this section are evaluated for a stationary Maxwellian ion distribution,

\begin{equation}\label{eq:maxwellian}
  f_M(\mathbf{v}) = \left(\frac{m_i}{2\pi kT_i}\right)^{3/2}
    \exp\!\left(-\frac{m_i |\mathbf{v}|^2}{2\,kT_i}\right),
\end{equation}

with $kT_i=1$\,keV.  Sec.~\ref{sec:rotation} integrates the same maps against a rotating, shifted Maxwellian.  The Maxwellian expectation over the domain $\Omega$ of Sec.~\ref{sec:loss_cone} is defined as

\begin{equation}\label{eq:expect}
  \langle h \rangle \;\equiv\;
    \frac{\int h(\KE,\eta)\,f_M(\mathbf{v})\,d^3v}{\int f_M(\mathbf{v})\,d^3v}
    \;=\;
    \frac{1}{Z_\text{grid}} \int_\Omega
    h(\KE,\eta)\,\IOLweight\;d\eta\,d\KE,
\end{equation}

where the velocity integrals extend over all gyrophases and over the region of velocity space with $(\KE,\eta)\in\Omega$, and $h(\KE,\eta)$ is any dimensionless test function.  The second form follows from $d^3v = v^2\,dv\,\sin\eta\,d\eta\,d\varphi$ with gyrophase $\varphi$ and $v=\sqrt{2\KE/m_i}$.  The normalization is

\begin{equation}\label{eq:Zgrid}
  Z_\text{grid} = \int_\Omega \IOLweight\;d\eta\,d\KE.
\end{equation}

Starting from a Maxwellian ion distribution and changing variables from velocity to $(\KE,\eta)$, the three Bayesian dimensionless transport moments (analogs of those in Refs.~\cite{stacey2011,stacey2015,wilks2016,piper2019}) are

\begin{align}
  \widetilde{\mathcal{N}}_\text{loss} &= \bigl\langle P(\text{lost}\mid\KE,\eta) \bigr\rangle,
  \label{eq:Nloss}\\
  \widetilde{\mathcal{E}}_\text{loss} &= \frac{1}{\tfrac{3}{2}kT_i}\bigl\langle P(\text{lost}\mid\KE,\eta)\,\KE \bigr\rangle,
  \label{eq:Eloss}\\
  \widetilde{\mathcal{M}}_\text{loss} &= \frac{1}{\sqrt{kT_i}}\Bigl\langle P(\text{lost}\mid\KE,\eta)\,\sqrt{\KE}\,\cos\eta \Bigr\rangle,
  \label{eq:Mloss}
\end{align}

where $\sqrt{\KE/kT_i}\,\cos\eta = v_\parallel/v_\text{th}$ with $v_\text{th}=\sqrt{2kT_i/m_i}$.  

In the Bayesian formulation, posterior uncertainty propagates approximately to each normalized moment.  Under the Laplace approximation $w\sim\mathcal{N}(w_\text{MAP},\Sigma)$, each moment $\widetilde{\mathcal{I}}$ is a smooth functional of $w$, and a first-order delta-method approximation gives $\mathrm{Var}[\widetilde{\mathcal{I}}] \approx \mathbf{d}^\top \Sigma\, \mathbf{d}$, where the gradient vector is

\begin{equation}\label{eq:delta_method}
    d_k = \bigl\langle P(\text{lost}\mid\KE,\eta)\,\bigl(1-P(\text{lost}\mid\KE,\eta)\bigr)\,u_k\,q \bigr\rangle,
\end{equation}

where $q(\KE,\eta)$ is the moment kernel ($1$, $\KE/(\tfrac{3}{2}kT_i)$, or $\sqrt{\KE/kT_i}\,\cos\eta$ for $\widetilde{\mathcal{N}}$, $\widetilde{\mathcal{E}}$, $\widetilde{\mathcal{M}}$ respectively) and $\Sigma$ is the Laplace covariance of the fitted probabilistic model (Eq.~\eqref{eq:laplace}).  Both $\Sigma$ and $u_k$ are available from the fitted model, so no additional GC simulations are required.  This propagated uncertainty is not available from a binary-label database, for which integration against $f_M$ yields a Monte Carlo average with no model-based uncertainty estimate.

\begin{figure}[!htbp]
  \centering
  \includegraphics[width=\linewidth,keepaspectratio]{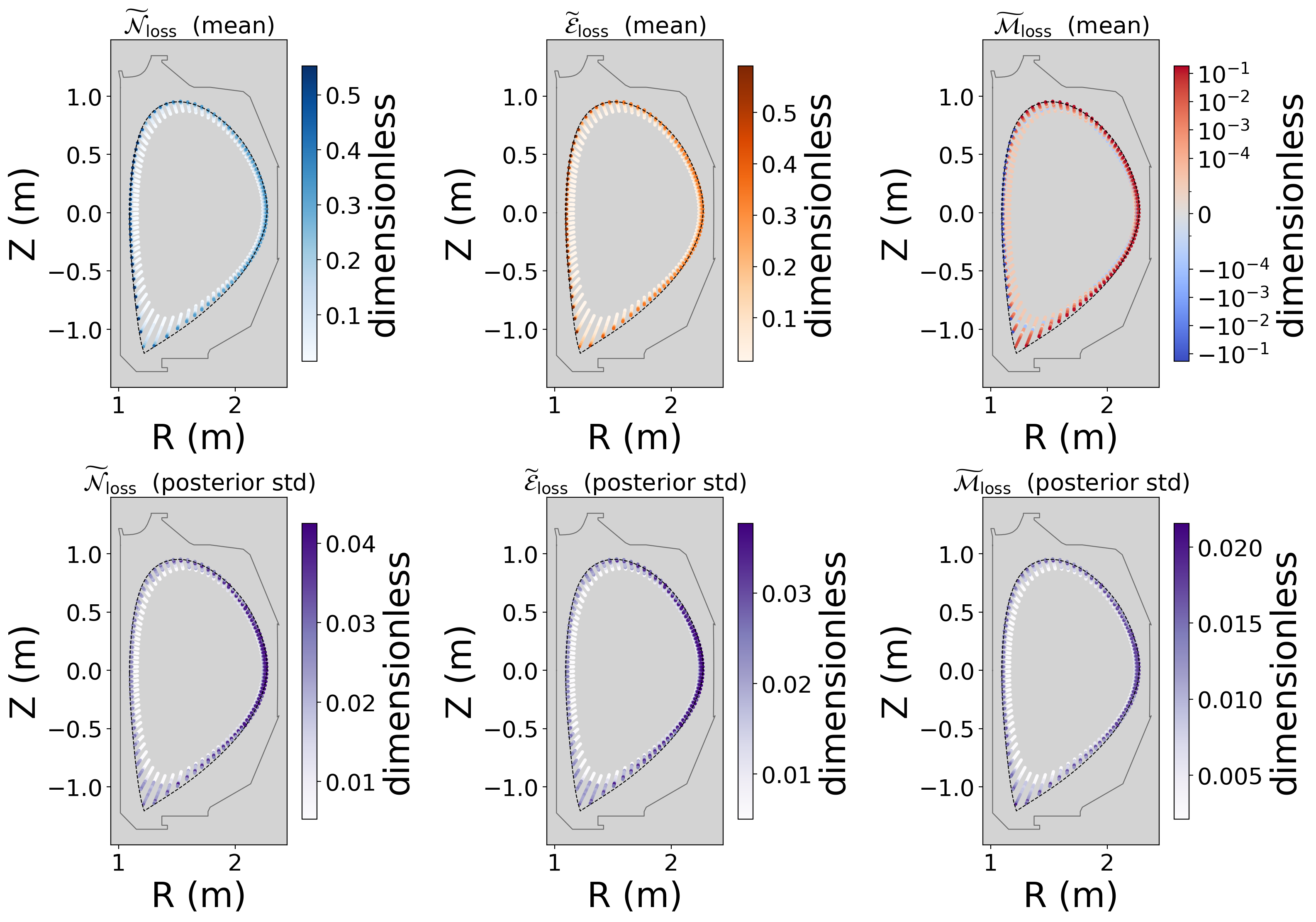}
  \caption{Spatial distribution of the three dimensionless IOL transport moments over the launch-position mesh.  Each column corresponds to one moment ($\widetilde{\mathcal{N}}_\text{loss}$, $\widetilde{\mathcal{E}}_\text{loss}$, $\widetilde{\mathcal{M}}_\text{loss}$); the upper row shows the MAP estimate and the lower row shows the posterior standard deviation from Eq.~\eqref{eq:delta_method}.  Scatter point color encodes the dimensionless value at each $(R,Z)$ launch position; the dashed contour marks the separatrix.  The mean $\widetilde{\mathcal{M}}_\text{loss}$ panel uses a symmetric logarithmic color scale to resolve both the sign and the wide dynamic range of the parallel-momentum loss.  All three quantities peak near the separatrix and on the outboard midplane, while posterior uncertainty is largest near the loss-cone boundary where the classifier is least certain.}
  \label{fig:iol_rz_maps}
\end{figure}

\begin{figure}[!htbp]
  \centering
  \begin{subfigure}[t]{\linewidth}
    \centering
    \includegraphics[width=\linewidth,keepaspectratio]{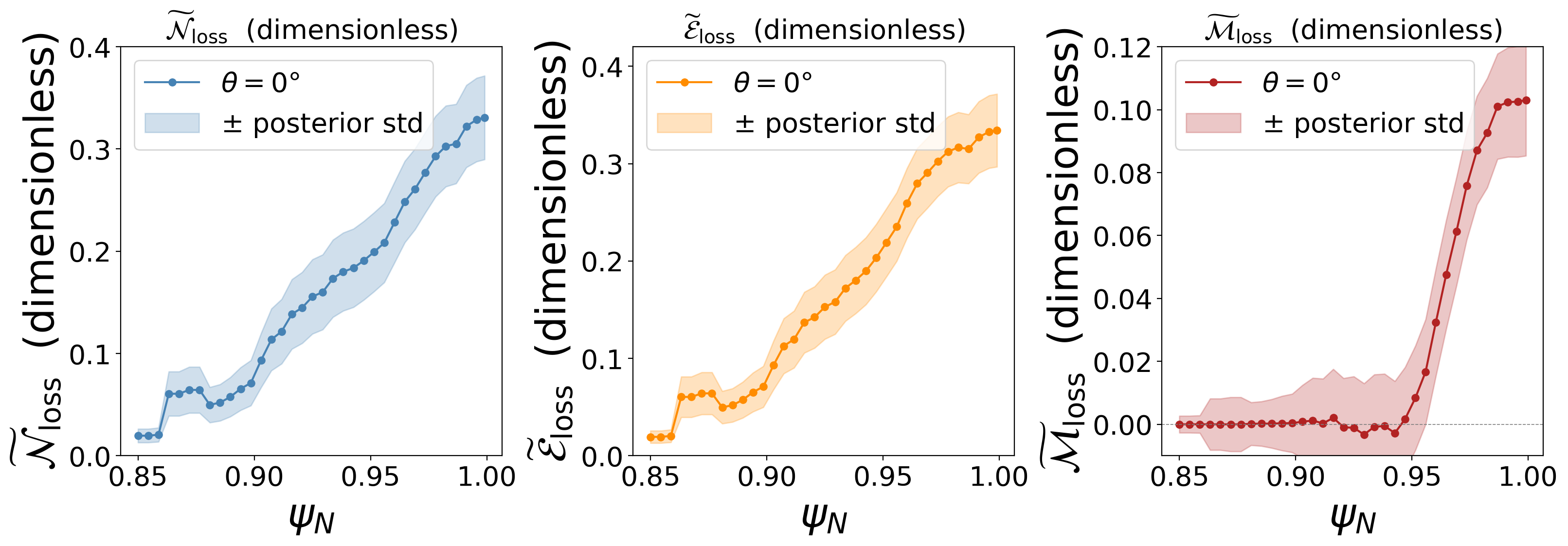}
    \caption{Radial profile at the outboard midplane ($\theta=0^\circ$).  Each marker shows the MAP normalized moment at that flux surface; the shaded band is the $\pm1\sigma$ posterior uncertainty from Eq.~\eqref{eq:delta_method}.}
    \label{fig:iol_psin_slice}
  \end{subfigure}
  \par\medskip
  \begin{subfigure}[t]{\linewidth}
    \centering
    \includegraphics[width=\linewidth,keepaspectratio]{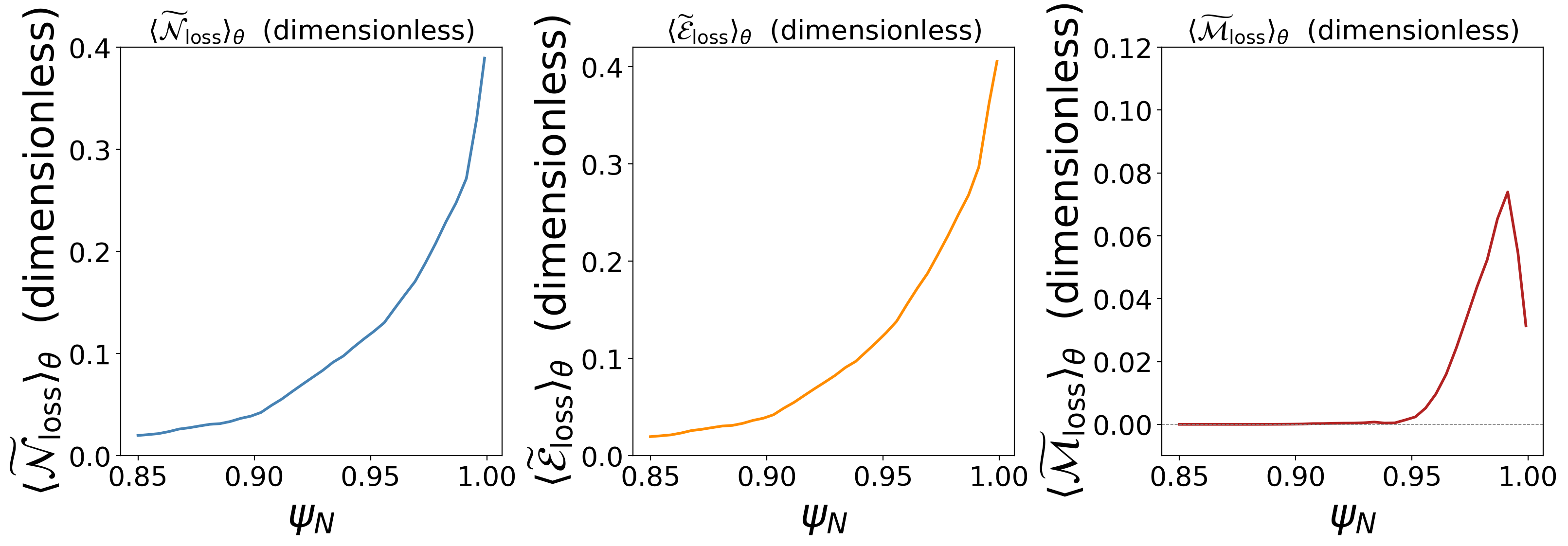}
    \caption{Radial profile averaged over all 100 poloidal angles, $\langle I\rangle_\theta(\psi_N)=\frac{1}{2\pi}\int_0^{2\pi} I(\psi_N,\theta)\,d\theta$, evaluated by periodic trapezoidal quadrature on the uniform $\theta$ grid.  No uncertainty band is shown because the per-position models are fit independently and do not provide the cross-angle covariance required for rigorous uncertainty propagation of the angle-averaged profile.}
    \label{fig:iol_psin_integrated}
  \end{subfigure}
  \caption{Radial profiles of the three dimensionless IOL transport moments ($\widetilde{\mathcal{N}}_\text{loss}$, $\widetilde{\mathcal{E}}_\text{loss}$, $\widetilde{\mathcal{M}}_\text{loss}$) as functions of normalized poloidal flux $\psi_N$.  Frame~(a) shows the single-angle slice at the outboard midplane; frame~(b) shows the corresponding poloidal-angle average.  All three quantities grow steeply as $\psi_N$ approaches 1, consistent with prior IOL calculations showing stronger edge losses near the separatrix \cite{stacey2011}.  The parallel-momentum loss $\widetilde{\mathcal{M}}_\text{loss}$ remains predominantly positive in the poloidal-angle-averaged profile and across the outboard-edge slice, apart from near-zero excursions at the innermost surfaces.}
  \label{fig:iol_psin_profiles}
\end{figure}

Evaluating Eqs.~\eqref{eq:Nloss}--\eqref{eq:Mloss} across the launch-position database using Eq.~\eqref{eq:delta_method} for the posterior standard deviation produces the spatial maps shown in Fig.~\ref{fig:iol_rz_maps} and the radial profiles in Fig.~\ref{fig:iol_psin_profiles}.  The spatial maps confirm that all three dimensionless moments are concentrated in the high-$\psi_N$ edge layer and are largest on the outboard midplane, where the loss-cone area in velocity space is greatest.  The posterior standard deviation tracks the loss-cone boundary geometry.  Positions near the separatrix show larger absolute uncertainty because the boundary cuts across a steeper sigmoid transition, while lower-$\psi_N$ edge positions carry negligible uncertainty.  At the 881 positions where all 785 GC trajectories are confined, all with $\psi_N\le0.96$, the fitted model still gives $\widetilde{\mathcal{N}}_\text{loss}=0.015$, which sets the floor of the maps at low $\psi_N$.  The $\widetilde{\mathcal{M}}_\text{loss}$ map changes sign across the full poloidal sweep.  It is positive at 90\% of the 1235 positions with $|\widetilde{\mathcal{M}}_\text{loss}|>10^{-3}$, so lost ions carry net parallel momentum along $\mathbf{B}$.  The $\psi_N$ profiles in Fig.~\ref{fig:iol_psin_profiles} show the same characteristic steep rise toward the separatrix seen in prior IOL studies \cite{stacey2011,piper2019}, indicating that the Bayesian moments reproduce the expected edge-localized trend while providing approximate posterior uncertainty at each launch position.

Fig.~\ref{fig:iol_psin_profiles}(a) shows per-position posterior uncertainty bands, but Fig.~\ref{fig:iol_psin_profiles}(b) does not.  The reason is that the poloidal-angle average combines moments from 100 independently fit Bayesian models, each providing only its own marginal variance.  Propagating uncertainty through the weighted average requires the full covariance between moments at different poloidal angles on the same flux surface, which the independent per-position fits do not supply.  Ion orbits couple different poloidal launch angles, because an ion launched at one $\theta$ passes through regions near other $\theta$ values.  The cross-angle covariances therefore need not be small and may be predominantly positive.  A rigorous uncertainty band for the angle-averaged profile would therefore require a joint probabilistic model across $\theta$, or at least a shared spatial prior over all positions on a fixed flux surface; this extension is left for future work.

\section{Edge Rotation from Orbit-Loss Torque}\label{sec:rotation}

Ions lost from the edge take their momentum with them.  Because more ions are lost in one toroidal direction than the other (Sec.~\ref{sec:integrals}), the loss removes net toroidal momentum from the edge plasma, which is expected to spin up in the opposite toroidal direction \cite{stacey2016}.  Orbit loss is one of several drives for edge rotation.  Turbulent diffusion acting on passing ions with finite orbit shifts produces a residual stress that provides a separate route to edge spin-up \cite{stoltzfusdueck2012}.  As an example of how the loss-probability database can feed a transport calculation, the orbit-loss torque is balanced against radial momentum diffusion across the flux surfaces of the launch mesh (Sec.~\ref{sec:results}).  Beyond the loss probabilities, the calculation needs the rate at which the loss region is refilled by new ions and the momentum diffusivity.  Neither is determined by the orbit calculations, so both are prescribed over a range of plausible values.  The loss probabilities are held fixed at $\mathbf{E}=0$, as in the database.  The rotation profiles below illustrate the edge response under these stated assumptions.

For toroidal rotation with an angular frequency $\omega(\psi_N)$ that is constant on each flux surface, the loss-weighted toroidal angular momentum per background ion at a launch position is
\begin{equation}\label{eq:ell_lost}
  \ell_\text{lost}(\psi_N,\theta;\omega) =
  \frac{\int P(\text{lost}\mid\KE,\eta)\, f_M(\mathbf{v}-\omega R\hat{\boldsymbol{\phi}})\, m_i R\, v_\phi\; d^3v}
       {\int f_M(\mathbf{v}-\omega R\hat{\boldsymbol{\phi}})\; d^3v},
\end{equation}
where $f_M$ is the Maxwellian of Eq.~\eqref{eq:maxwellian} shifted by the rotation, and $v_\phi$ is the toroidal component of the ion velocity at launch.  The denominator is the full background density, so $\ell_\text{lost}$ is an average over all background ions, and it is evaluated at the launch position.  In the present equilibrium $B_\phi<0$ while the plasma current is along $+\hat{\boldsymbol{\phi}}$, so the predominantly positive $\widetilde{\mathcal{M}}_\text{loss}$ of Fig.~\ref{fig:iol_psin_profiles} corresponds to counter-current loss and a negative $\ell_\text{lost}$.

Flux-surface averages are taken over the closed surfaces labeled by $\psi_N$, $\langle A\rangle_\psi = \oint A\,dl_p/B_p \big/ \oint dl_p/B_p$, with the volume derivative $V' = dV/d\psi_N = 2\pi|\Psi_\text{sep}-\Psi_\text{axis}|\oint dl_p/B_p$ and the dimensionless metric $G = \langle R^2|\nabla\psi_N|^2\rangle_\psi$.  For uniform density $n_i$, a diffusive angular-momentum flux $-m_i n_i D_\phi R^2\nabla\omega$, and replacement ions that carry no toroidal angular momentum, the steady balance is
\begin{equation}\label{eq:rotation_balance}
  \frac{1}{V'}\frac{d}{d\psi_N}\left[V' m_i n_i D_\phi\, G\,\frac{d\omega}{d\psi_N}\right]
  = n_i\,\nu_\text{eff}\,\bigl\langle \ell_\text{lost}(\omega)\bigr\rangle_\psi,
\end{equation}
where $D_\phi$ is the momentum diffusivity in $\mathrm{m^2/s}$ and $\nu_\text{eff}$ is the renewal rate of the loss region in $\mathrm{s^{-1}}$.  The applied torque density is $-n_i\nu_\text{eff}\langle\ell_\text{lost}\rangle_\psi$, so a negative $\langle\ell_\text{lost}\rangle_\psi$ spins the plasma up in the $+\hat{\boldsymbol{\phi}}$ direction.  The same conservative flux-coordinate structure underlies neoclassical and experimental momentum-transport analyses \cite{hinton1985neoclassical,chrystal2017predicting}.  The boundary conditions are $d\omega/d\psi_N=0$ at $\psi_N=0.85$ and $\omega=0$ at $\psi_N=0.999$.  The quantities $G$ and $V'$ are evaluated on high-resolution $\psi_N$ contours of the equilibrium.

\begin{figure}[!htbp]
  \centering
  \begin{subfigure}[t]{0.32\linewidth}
    \centering
    \includegraphics[width=\linewidth]{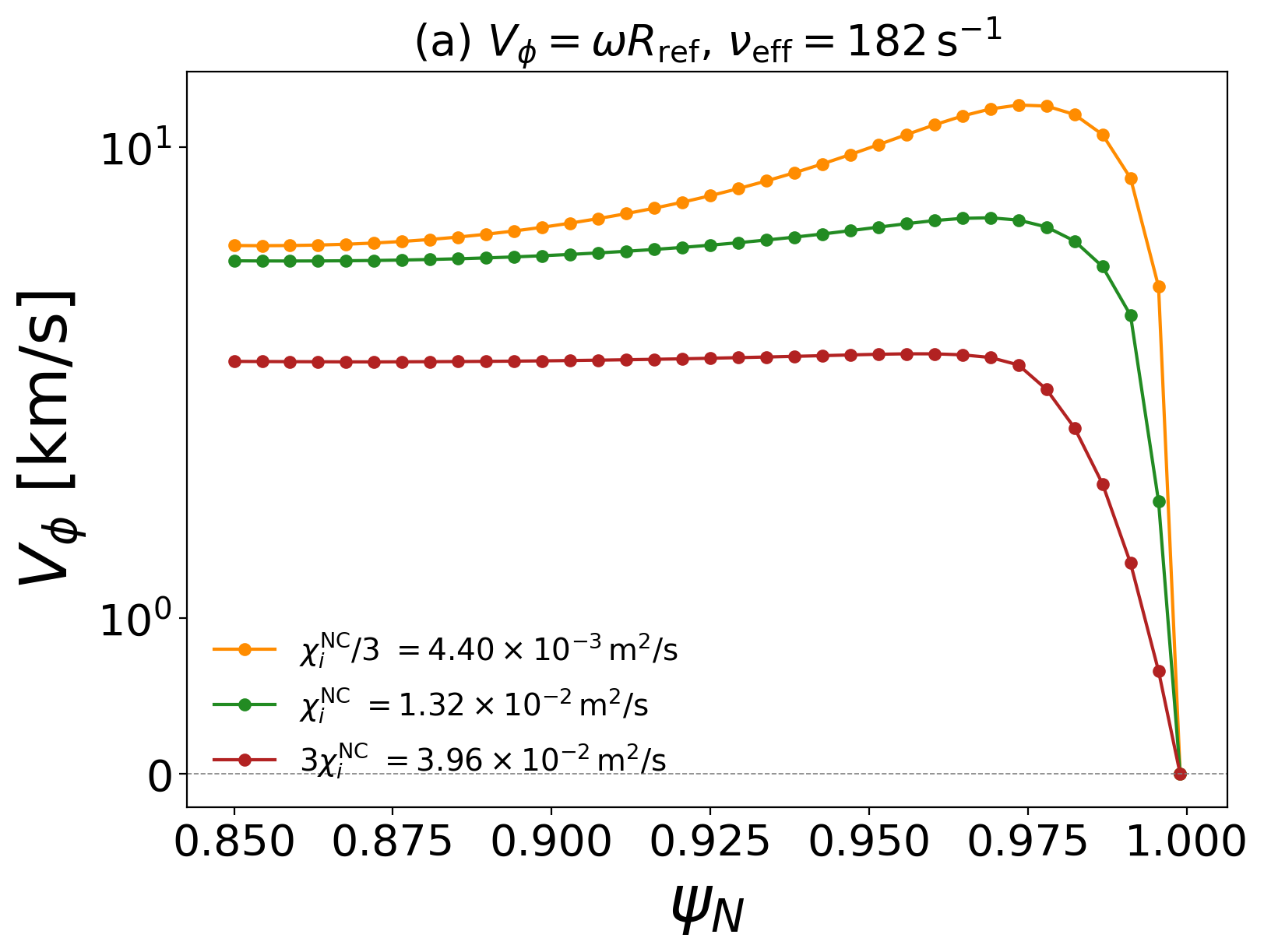}
  \end{subfigure}
  \hfill
  \begin{subfigure}[t]{0.32\linewidth}
    \centering
    \includegraphics[width=\linewidth]{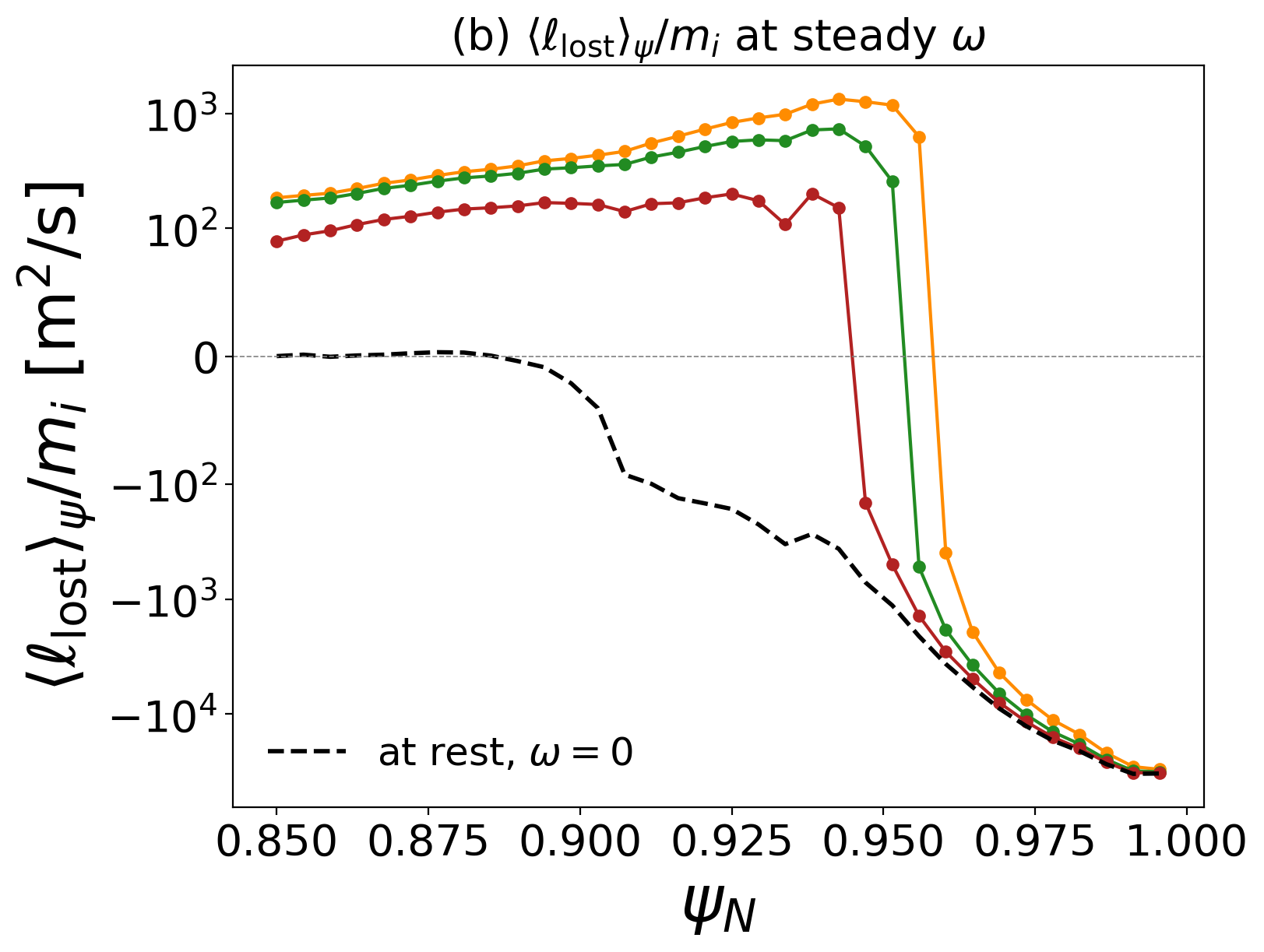}
  \end{subfigure}
  \hfill
  \begin{subfigure}[t]{0.32\linewidth}
    \centering
    \includegraphics[width=\linewidth]{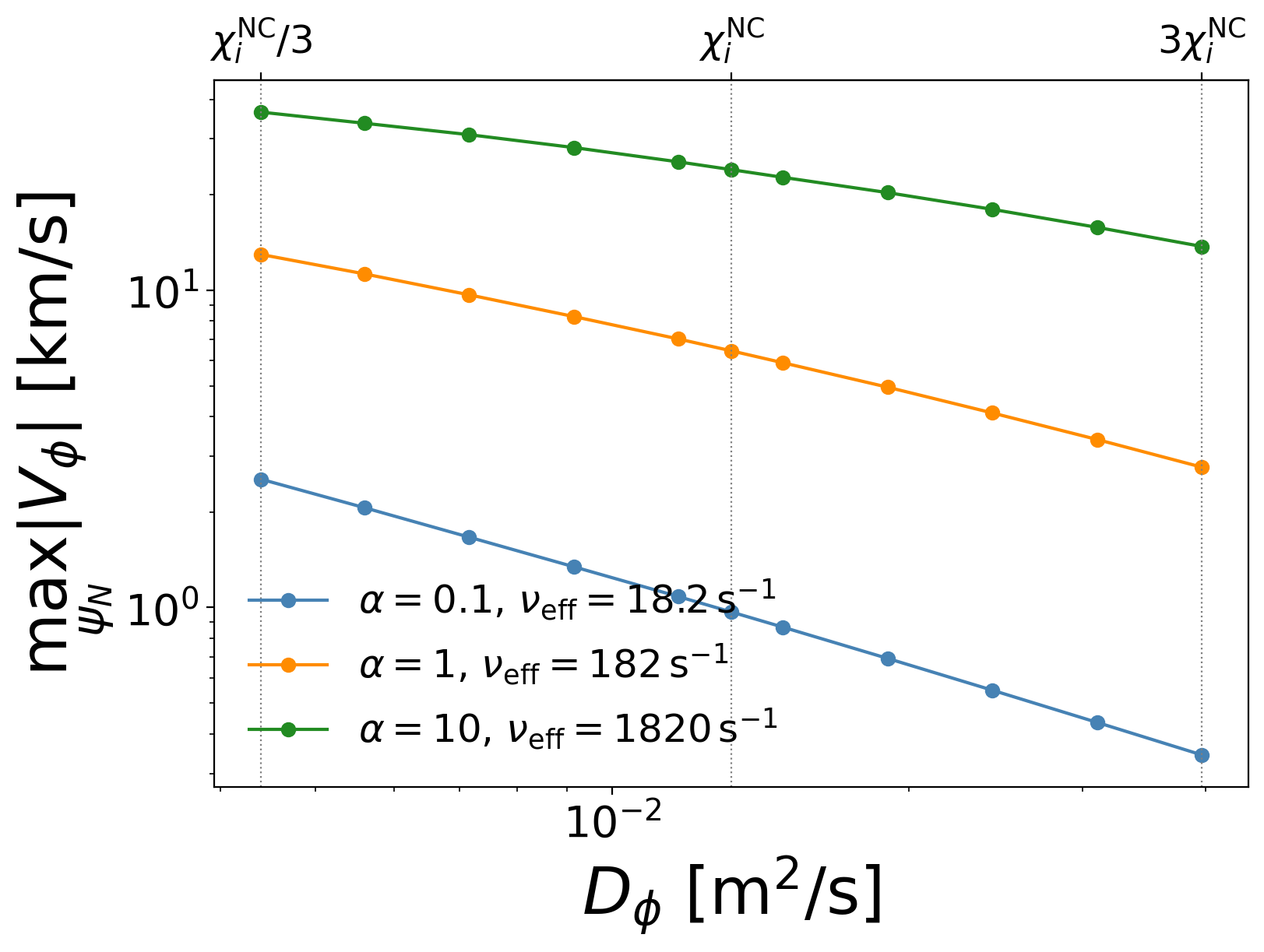}
  \end{subfigure}
  \caption{Steady rotation from Eq.~\eqref{eq:rotation_balance} for the analytic equilibrium and deuterium at $kT_i=1$\,keV.  Frames (a) and (b) use renewal rate $\nu_\text{eff}=\nu_i=182\,\mathrm{s^{-1}}$ ($\alpha=1$); the labels in frame (a) identify the three heat-linked momentum diffusivities used in both panels.  Markers in frames (a) and (b) correspond to the sampled flux surfaces.  Frame (a) shows the toroidal velocity $V_\phi=\omega R_\text{ref}$ with $R_\text{ref}=\langle R\rangle_\psi$, which vanishes at the outer boundary $\psi_N=0.999$.  Frame (b) shows the flux-surface-averaged lost angular momentum per unit ion mass, $\langle\ell_\text{lost}\rangle_\psi/m_i$, at the steady rotation of each profile in frame (a) and at rest (black dashed curve); the outermost surface, which carries the boundary condition, is omitted.  Frame (c) shows peak toroidal velocity $\max_{\psi_N}|V_\phi|$ as a function of constant momentum diffusivity $D_\phi$ for $\nu_\text{eff}=\alpha\nu_i$ with $\alpha=0.1$, 1, and 10; the 11 diffusivities span $\chi_i^\text{NC}/3$ to $3\chi_i^\text{NC}$, and the three anchors are marked by vertical dotted lines and labeled on the upper axis.  Both vertical axes in frames (a) and (b) use a symmetric logarithmic scale, and the horizontal dashed line marks zero.  The loss probabilities are held fixed at $\mathbf{E}=0$.}
  \label{fig:rotation_profiles}
\end{figure}

The momentum diffusivity $D_\phi$ and the renewal rate $\nu_\text{eff}$ are set from neoclassical estimates evaluated at the outboard point of the $\psi_N=0.97$ surface.  For deuterium at $kT_i=1$\,keV, with an assumed density $n_i=10^{19}\,\mathrm{m^{-3}}$ and Coulomb logarithm $\ln\Lambda=17$, the Braginskii ion collision rate is $\nu_i=182\,\mathrm{s^{-1}}$.  The surface has safety factor $q=3.82$ and local inverse aspect ratio $\epsilon=0.28$.  The ion collisionality $\nu_*=\nu_i qR_0/(\epsilon^{3/2}v_\text{th})$ is the ratio of the effective collision frequency of trapped ions to their bounce frequency.  Its value $\nu_*=0.03$ lies in the banana regime, $\nu_*<1$.  The banana-regime, large-aspect-ratio results of Hinton and Wong \cite{hinton1985neoclassical} for the momentum diffusivity and the ion heat diffusivity,
\begin{align}
  D_\phi^\text{NC} &= 0.1\,\epsilon^2\rho_{i\theta}^2\,\nu_i, \label{eq:D_nc}\\
  \chi_i^\text{NC} &= 0.66\,\epsilon^{1/2}\rho_{i\theta}^2\,\nu_i, \label{eq:chi_nc}
\end{align}
give $D_\phi^\text{NC}=2.98\times10^{-4}\,\mathrm{m^2/s}$ and the ion heat diffusivity $\chi_i^\text{NC}=1.32\times10^{-2}\,\mathrm{m^2/s}$ for $\epsilon=0.28$ and the poloidal gyroradius $\rho_{i\theta}=m_i v_\text{th}/(eB_\theta)=1.44\,\mathrm{cm}$.  The Prandtl number $\mathrm{Pr}=D_\phi/\chi_i$ is thus $0.02$ for the neoclassical estimates, with $D_\phi^\text{NC}$ a factor of 44 below $\chi_i^\text{NC}$.  Measured momentum diffusivities are typically comparable to the ion heat diffusivity \cite{degrassie2009tokamak}, so $D_\phi$ is set from $\chi_i^\text{NC}$ with Prandtl numbers $\mathrm{Pr}=1/3$, 1, and 3.  These three diffusivities set the profiles of Fig.~\ref{fig:rotation_profiles}(a) and \ref{fig:rotation_profiles}(b).  For the peak-velocity scan of Fig.~\ref{fig:rotation_profiles}(c), eight logarithmically spaced intermediate values are added, giving 11 constant diffusivities between $\chi_i^\text{NC}/3$ and $3\chi_i^\text{NC}$.  The renewal rate is modeled as $\nu_\text{eff}=\alpha\nu_i$ with $\alpha=0.1$, 1, and 10, so the scan covers 33 scenarios.  This relaxation model assumes that ions escape rapidly once they enter the loss region; the actual supply is set by pitch-angle scattering and streaming into the loss-cone \cite{degrassie2015thermal,zhu2023}, which the probability maps alone do not determine.  The density cancels from Eq.~\eqref{eq:rotation_balance} and enters only through $\nu_i$ and the coefficient estimates.

Eq.~\eqref{eq:rotation_balance} is discretized with a conservative finite-volume operator on the 35 flux surfaces of the database, and $\langle\ell_\text{lost}\rangle_\psi$ is evaluated from all 3500 Bayesian probability maps on a $100\times100$ $(\KE,\eta)$ grid.  Because the source depends on $\omega$ through the shifted Maxwellian, the nonlinear system is solved by a damped Newton iteration starting from $\omega=0$.  The largest toroidal rotation speed reaches $0.12\,v_\text{th}$.

Fig.~\ref{fig:rotation_profiles} shows the rotation profiles and lost angular momentum for the three anchors at $\alpha=1$.  All three profiles rotate in the $+\hat{\boldsymbol{\phi}}$ direction, co-current with the plasma current, consistent with orbit-loss calculations of co-current edge spin-up \cite{pan2014cocurrent,zhu2023}.  The profile shape and amplitude are set by $D_\phi$.  At $3\chi_i^\text{NC}$ the profile is nearly flat at $2.6$ to $2.8$\,km/s out to $\psi_N\approx0.96$ and then falls to zero at the outer boundary.  At $\chi_i^\text{NC}/3$ the rotation rises from $5.4$\,km/s at $\psi_N=0.85$ to a peak of $13$\,km/s at $\psi_N=0.97$.  Fig.~\ref{fig:rotation_profiles}(b) shows that the rotation feeds back on its own source.  At rest, $\langle\ell_\text{lost}\rangle_\psi$ is negative on every surface from $\psi_N=0.89$ outward.  At the steady rotation of Eq.~\eqref{eq:rotation_balance}, the shifted Maxwellian makes it positive inside $\psi_N\approx0.95$ to $0.96$ and weakens it slightly near the edge.  At $\psi_N=0.996$ for $\chi_i^\text{NC}/3$, $\langle\ell_\text{lost}\rangle_\psi/m_i$ changes from $-3.3\times10^4$ to $-3.0\times10^4\,\mathrm{m^2/s}$.  The torque is therefore concentrated on the outermost surfaces.  The net applied torque at the assumed density ranges from $0.01$ to $0.66$\,N\,m across the 33 scenarios.

Fig.~\ref{fig:rotation_profiles}(c) summarizes the full scenario matrix through the peak velocity.  The peak decreases monotonically with $D_\phi$ and increases with $\alpha$, spanning $0.34$ to $2.5$\,km/s for $\alpha=0.1$, $2.8$ to $13$\,km/s for $\alpha=1$, and $14$ to $36$\,km/s for $\alpha=10$.  The response to renewal weakens at lower diffusivity: successive tenfold increases of $\alpha$ raise the peak by factors of $8.1$ and $5.0$ at $3\chi_i^\text{NC}$, and by $5.1$ and $2.8$ at $\chi_i^\text{NC}/3$.  This weakening is consistent with the source feedback of Fig.~\ref{fig:rotation_profiles}(b).  At fixed $\alpha$, moving from $3\chi_i^\text{NC}$ to $\chi_i^\text{NC}/3$ raises the peak by a factor of $2.7$ to $7.4$, so the rotation amplitude depends on the prescribed transport and renewal coefficients at least as strongly as on the loss probabilities.

\section{Neural Network Surrogate}\label{sec:NN}

The Bayesian probability maps developed in Sec.~\ref{sec:results} provide the reference IOL predictions at each launch position, but the per-position Bayesian AMR workflow remains computationally expensive when full coverage across the poloidal mesh is required.  To accelerate repeated in-domain queries on the same launch-position mesh, we train a neural network surrogate that learns the mapping from launch coordinates to loss-probability images.

\subsection{Architecture and Training}

The surrogate maps normalized launch coordinates $(\psi_N, \theta)$ of Sec.~\ref{sec:results} to a probability image $P(\text{lost}\mid \KE, \eta)$ discretized on a $64 \times 64$ grid spanning the domain $\Omega$ of Sec.~\ref{sec:loss_cone}.  The sampled edge interval is rescaled to $\tilde{\psi}_N = (\psi_N - \psi_{N,\min})/(\psi_{N,\max} - \psi_{N,\min}) \in [0,1]$, where $\psi_{N,\min}=0.85$ and $\psi_{N,\max}=0.999$ are the mesh bounds of Sec.~\ref{sec:database_setup}.  This rescaled $\tilde\psi_N$ is distinct from $\psi_N$ itself, which is normalized over the full device; $\tilde\psi_N$ is retained directly as a network input, while the poloidal angle $\theta$ is mapped to its cyclic representation $(\sin\theta, \cos\theta)$, yielding a three-dimensional input vector $\mathbf{x} = (\tilde{\psi}_N, \sin\theta, \cos\theta)$.

The architecture design is motivated by two properties of the IOL probability maps.  First, the loss-cone boundary represents a sharp transition from $P\approx0$ to $P\approx1$ over a narrow band in $(\KE,\eta)$ space; standard multi-layer perceptrons (MLPs) exhibit spectral bias toward low-frequency functions and struggle to represent such sharp transitions without prohibitive depth \cite{rahaman2019spectral}.  Second, the output is a spatially structured image in which neighboring pixels are correlated.  The probability field varies smoothly away from the boundary, and the boundary itself is a continuous curve whose shape changes gradually with launch position.  These considerations lead to an encoder-decoder architecture with Fourier feature encoding at the input stage.  The input coordinates are projected through a random Fourier feature layer \cite{tancik2020fourier}:
\begin{equation}\label{eq:fourier_features}
  \gamma(\mathbf{x}) = \left[\mathbf{x},\, \sin(2\pi \mathbf{B}\mathbf{x}),\, \cos(2\pi \mathbf{B}\mathbf{x})\right],
\end{equation}
where $\mathbf{B} \in \mathbb{R}^{N_f \times 3}$ is a fixed random matrix with entries drawn from $\mathcal{N}(0, \sigma_f^2)$ with $\sigma_f = 1$, and $N_f = 32$ is the number of Fourier frequencies.

The encoded features pass through an MLP encoder consisting of four residual blocks with hidden dimension 256, each containing layer normalization, GELU activation, and dropout regularization (rate 0.1).  The MLP output is projected to a spatial feature tensor and reshaped to an $8\times8$ grid with 256 channels.  A convolutional decoder then upsamples this representation through a sequence of bilinear upsampling and residual convolution blocks (three per stage, with base channel count 64) until the target $64\times64$ resolution is reached.  The final layer applies a sigmoid activation to produce probabilities in $[0,1]$.

Training uses the AdamW optimizer with a cosine-annealed learning rate schedule and linear warmup.  The loss function combines three terms:
\begin{equation}\label{eq:nn_loss}
  \mathcal{L} = \mathcal{L}_{\mathrm{MSE}} + \lambda_{\mathrm{grad}}\mathcal{L}_{\mathrm{grad}} + \lambda_{\mathrm{bdy}}\mathcal{L}_{\mathrm{bdy}},
\end{equation}
where $\mathcal{L}_{\mathrm{MSE}}$ is the pixel-wise mean squared error (MSE), $\mathcal{L}_{\mathrm{grad}}$ penalizes deviations in the spatial gradient field, and $\mathcal{L}_{\mathrm{bdy}}$ applies additional weight to pixels near the loss-cone boundary (where $P \approx 0.5$).  $\mathcal{L}_{\mathrm{grad}}$ is the mean squared error between the finite-difference gradients of the predicted and target images (first differences along each image axis), and $\mathcal{L}_{\mathrm{bdy}}$ is the squared pixel error weighted by $4P(1-P)$ evaluated on the target map.  This weight peaks at the $P=0.5$ boundary and vanishes in confidently classified regions.  The gradient term is motivated by the observation that the loss-cone boundary is characterized by a localized region of high $|\nabla P|$; preserving gradient structure keeps the predicted boundary sharp.  The boundary-weighted term compensates for the thinness of the boundary, which would otherwise contribute negligibly to the MSE.  The weighting coefficients $\lambda_{\mathrm{grad}} = 0.1$ and $\lambda_{\mathrm{bdy}} = 0.5$ were selected to balance these terms throughout training.  Training runs for up to 1500 epochs with batch size 64, an initial learning rate of $3\times10^{-5}$ (150-epoch linear warmup followed by cosine annealing), weight regularization coefficient of $5\times10^{-4}$, and gradient-norm clipping at 0.5; early stopping monitors the validation loss with a patience of 300 epochs.

The database of 3500 Bayesian probability maps is split into training (3100 samples), validation (300 samples), and test (100 samples) sets.  The validation and test sets are drawn exclusively from interior mesh positions, whose $(\psi_N, \theta)$ indices lie strictly inside the grid boundaries, while all boundary positions are retained in the training set.  The reported metrics therefore measure interpolation between nearby training positions on the sampled mesh.

\subsection{Prediction Performance}

\begin{figure}[!htbp]
  \centering
  \includegraphics[width=0.55\linewidth]{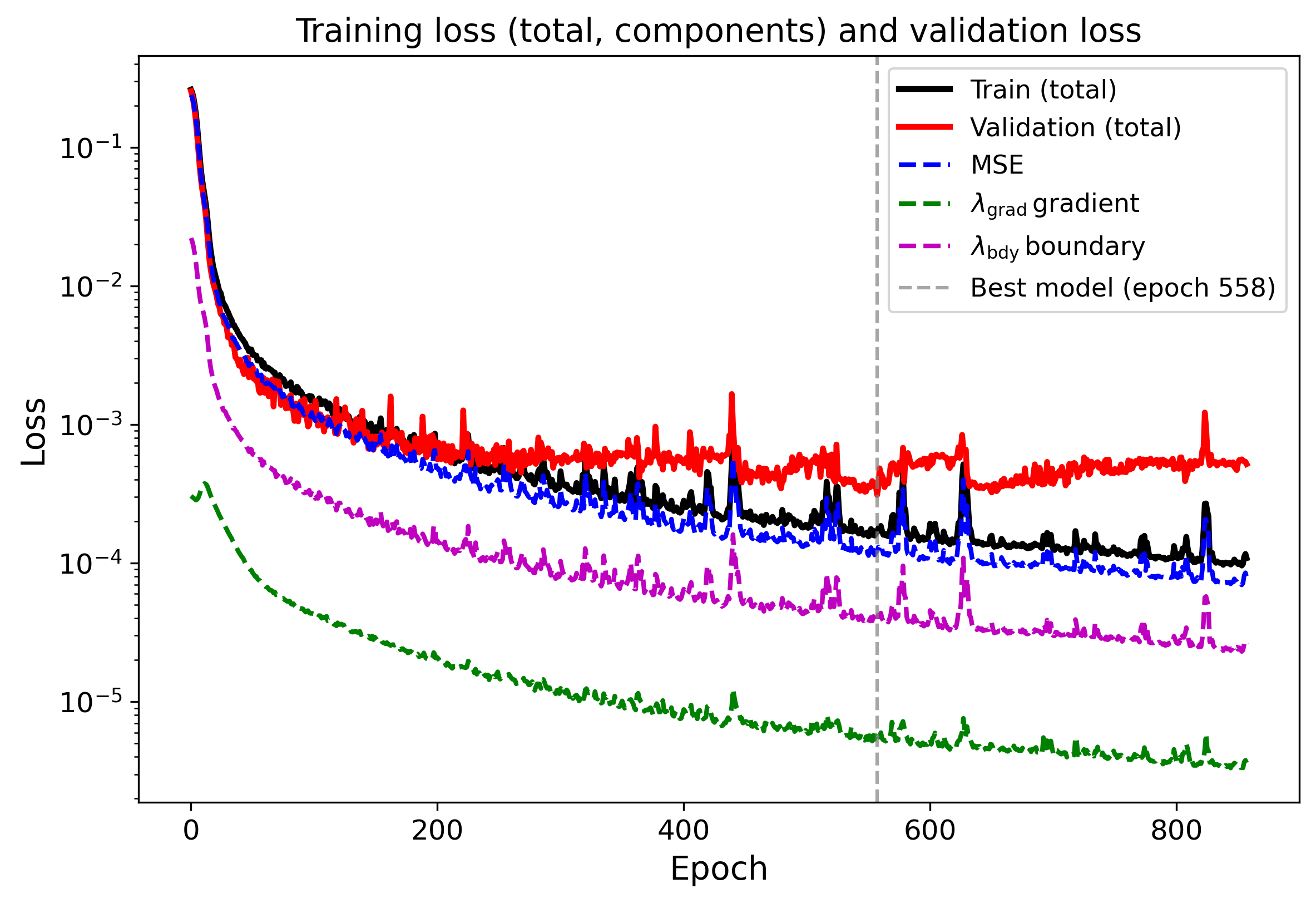}
  \caption{Total training loss, its weighted pixel-wise MSE, gradient, and boundary components from Eq.~\eqref{eq:nn_loss}, and total validation loss on a logarithmic scale.  All components decrease consistently, and the validation curve tracks the training curve until it flattens; the vertical dashed line marks the epoch at which the best model (minimum validation loss) was saved.}
  \label{fig:nn_training_curves}
\end{figure}

Fig.~\ref{fig:nn_training_curves} shows the evolution of the total training loss, its three weighted components from Eq.~\eqref{eq:nn_loss}, and the total validation loss.  All three components decrease by more than two orders of magnitude.  The validation loss closely tracks the training loss through the bulk of training and then flattens while the training loss continues to decrease; early stopping selects the model at the validation minimum, guarding against the mild overfitting that develops beyond it.  Table~\ref{tab:nn_metrics} reports the final prediction accuracy on each data split in terms of the MSE, the mean absolute error (MAE), and the root mean squared error (RMSE).  The validation and test MSE are $3.3$ and $4.2$ times the training MSE.

\begin{table}[t]
\centering
\caption{Neural network prediction accuracy on training, validation, and test sets.  All metrics are computed over the full $64\times64$ probability images.}\label{tab:nn_metrics}
\begin{tabular}{lccc}
\toprule
Split & MSE & MAE & RMSE \\
\midrule
Training    & $7.11 \times 10^{-5}$ & 0.0037 & 0.0084 \\
Validation  & $2.37 \times 10^{-4}$ & 0.0057 & 0.0154 \\
Test        & $3.01 \times 10^{-4}$ & 0.0057 & 0.0174 \\
\bottomrule
\end{tabular}
\end{table}

\begin{figure}[!htbp]
  \centering
  \includegraphics[width=\linewidth]{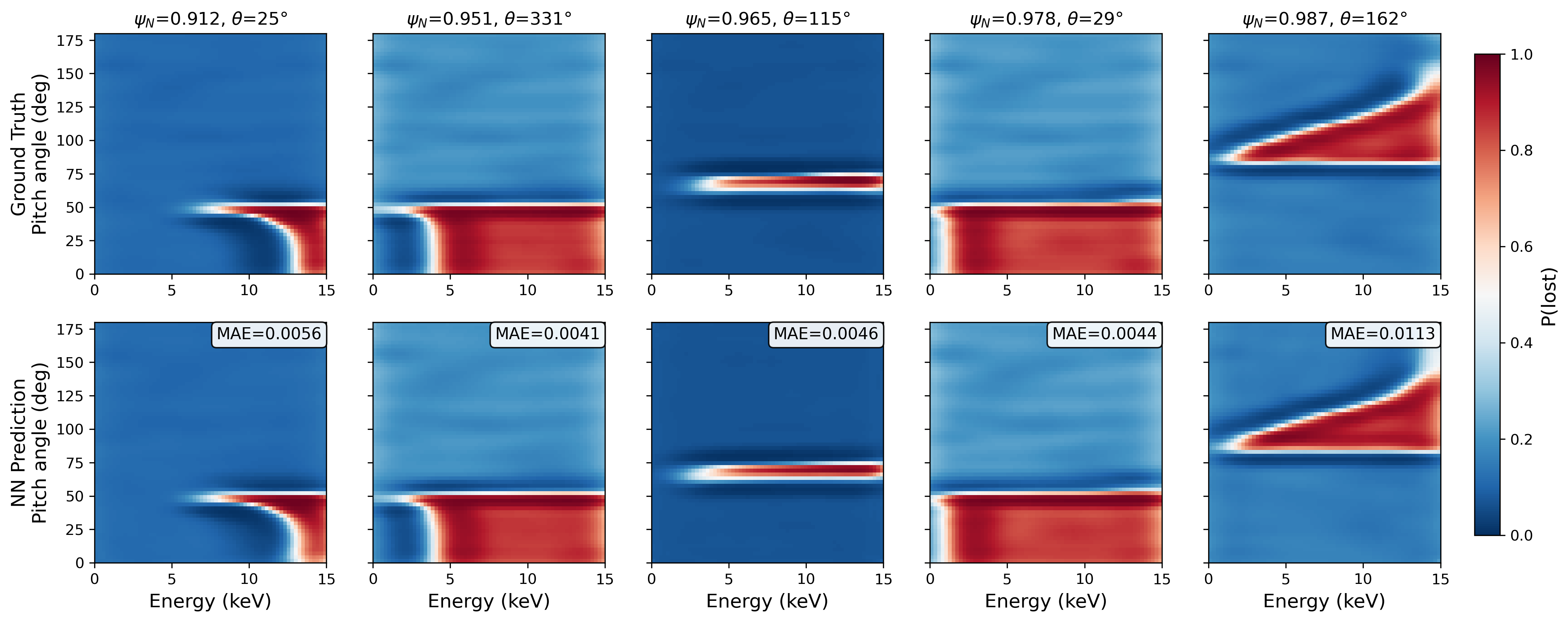}
  \caption{Comparison of ground-truth Bayesian probability maps (top row) and neural network predictions (bottom row) for five representative test-set positions.  Each column corresponds to a different launch position spanning the $(\psi_N, \theta)$ domain; the MAE is shown for each case.}
  \label{fig:nn_predictions}
\end{figure}

Fig.~\ref{fig:nn_predictions} compares ground-truth Bayesian probability maps with neural network predictions for five test-set launch positions across the sampled $(\psi_N, \theta)$ domain.  In three cases, loss is concentrated at low pitch angles and reaches lower energies as $\psi_N$ increases.  The other two show a narrow loss band near $\eta=70^\circ$ and a broad loss region at higher pitch angles.  The network reproduces these regions and their energy-dependent boundaries.

\subsection{Error Analysis and Sample Complexity}

\begin{figure}[!htbp]
  \centering
  \includegraphics[width=0.9\linewidth]{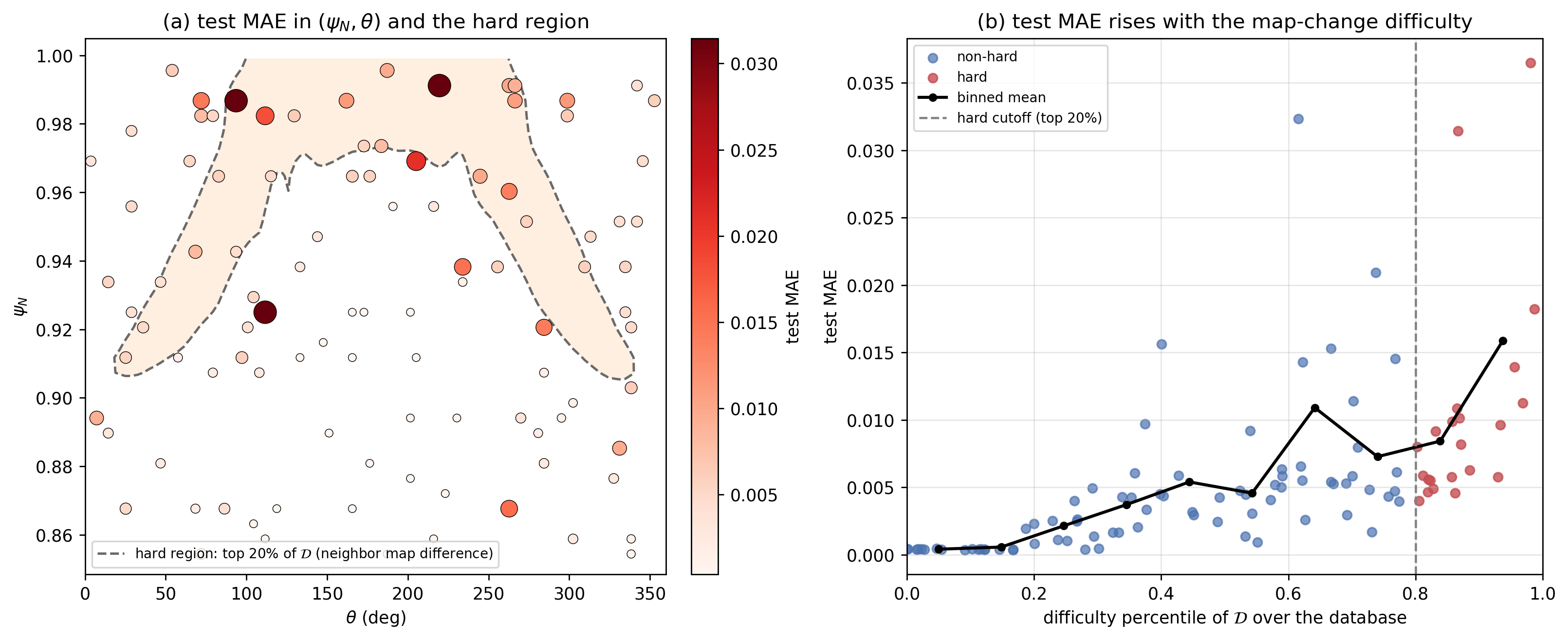}
  \caption{The neighbor-difference difficulty score $\mathcal{D}$ of Eq.~\eqref{eq:difficulty} predicts where the network errs.  (a) Test-position MAE at the $(\psi_N,\theta)$ launch positions, with marker color and size both scaling with the MAE, overlaid on the hard region (shaded, dashed boundary), which is the top $20\%$ of the database by the mesh-smoothed $\mathcal{D}$.  The hard region excludes quiescent high-$\psi_N$ sectors and extends radially inward along the two loss-cone transition bands near $\theta\approx90^\circ$ and $260^\circ$, matching the fast-transition structure visible in Fig.~\ref{fig:no_efield_grid}; the largest test errors fall inside it.  (b) Test MAE versus the percentile of $\mathcal{D}$ over the database; the dashed line marks the top-$20\%$ hard-region cutoff and the solid curve is the binned mean.  Test positions in the hard region have $2.4\times$ the mean MAE of the remaining test positions.}
  \label{fig:nn_hard_region_check}
\end{figure}

Fig.~\ref{fig:nn_hard_region_check}(a) maps the neural network prediction error over the $(\psi_N,\theta)$ launch domain, restricted to the 100 held-out test positions so that the pattern reflects genuine interpolation error.  The MAE increases toward the separatrix, where the loss-cone boundary occupies a larger fraction of velocity space and exhibits sharper structure.  The poloidal profile is peaked near $\theta \approx 90^\circ$ and $\theta \approx 260^\circ$, while the errors remain small around both midplanes, where the loss-cone maps, whether large (outboard) or nearly empty (inboard), vary only weakly between neighboring positions.  Inspection of the highest-error positions shows that in each case the target map contains an extended, sharp loss-cone boundary, as in the near-separatrix maps of the database overview in Fig.~\ref{fig:no_efield_grid}.  At the two poloidal peaks, the target map also differs strongly from the maps of the neighboring launch positions.  The residual error is concentrated in a thin band along that boundary.

The upper error peak at $\theta\approx90^\circ$ lies at the top of the plasma, diametrically opposite the lower X-point, so proximity to the X-point does not account for both peaks.  The error is organized by how rapidly the target probability maps change between neighboring launch positions.  This rate of change is quantified with a model-free, neighbor-difference difficulty score, defined for each launch position $i$ as
\begin{equation}\label{eq:difficulty}
  \mathcal{D}_i \;=\; \frac{1}{n_i}\sum_{j\,\in\,\mathrm{adj}(i)} \overline{\left|P_i - P_j\right|},
\end{equation}
where the sum runs over the $n_i$ mesh-adjacent launch positions of $i$ (the two $\theta$ neighbors on the same flux surface, treated periodically, and the $\psi_N$ neighbors on the adjacent surfaces), $P_i$ is the ground-truth Bayesian probability map, and $\overline{|P_i - P_j|}$ denotes the mean absolute difference over the $(\KE,\eta)$ grid.  $\mathcal{D}_i$ measures how different a position's loss-cone map is from the maps surrounding it, which is the variation a coordinate-to-image surrogate must bridge when interpolating.  To suppress mesh-scale noise, the $\mathcal{D}$ field is smoothed with a $3\times3$ moving average on the launch mesh (periodic in $\theta$), and the ``hard region'' is defined as the top $20\%$ of positions (700 of 3500) by the smoothed score.  Fig.~\ref{fig:nn_hard_region_check}(a) overlays this region on the test error.  The hard region excludes quiescent high-$\psi_N$ sectors where the loss-cone changes little between neighbors, and it extends radially inward along the two loss-cone transition bands, reproducing the fast-transition structure directly visible in the database overview of Fig.~\ref{fig:no_efield_grid}.  Fig.~\ref{fig:nn_hard_region_check}(b) shows the test MAE against the percentile of $\mathcal{D}$ over the database.  The error rises systematically with the score (rank correlation $0.80$), and test positions inside the hard region have a mean MAE $2.4\times$ larger than the remaining test positions ($0.0105$ versus $0.0044$).

\begin{figure}[!htbp]
  \centering
  \includegraphics[width=0.55\linewidth]{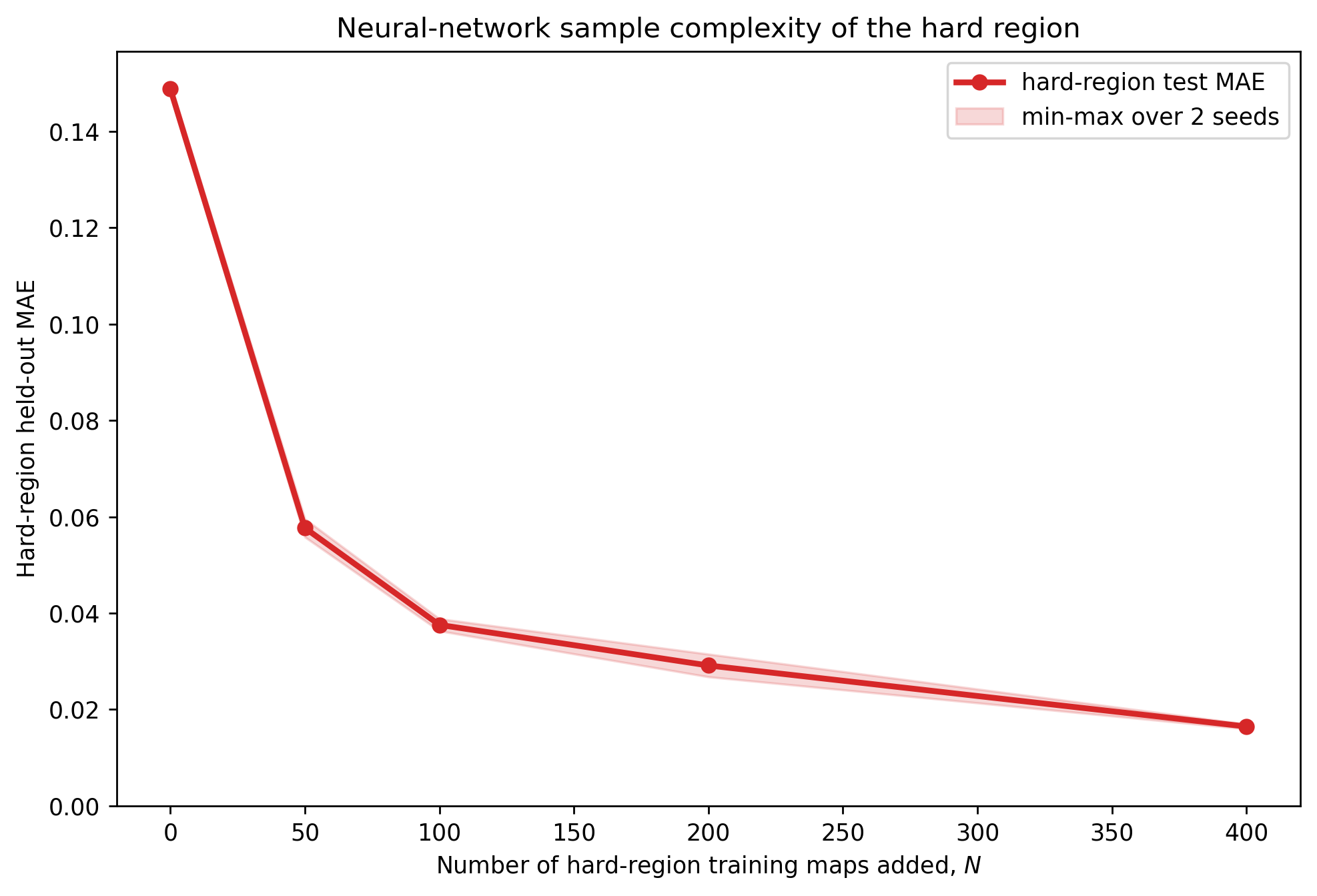}
  \caption{Sample complexity of the hard region.  The network is retrained from scratch on the fixed 2800 non-hard positions plus $N$ hard-region maps, and evaluated on 120 held-out hard-region positions; the shaded band spans two independent random draws per $N$ and is barely wider than the line, indicating the curve is robust to the particular maps drawn.  The held-out hard-region MAE drops by $61\%$ with the first 50 hard-region maps and by $89\%$ with 400.}
  \label{fig:nn_sample_complexity}
\end{figure}

The score $\mathcal{D}$ is computable for every position as soon as its own and its neighbors' Bayesian maps exist, before any surrogate is trained, so it can direct where the Bayesian workflow should generate additional maps.  To test whether the hard-region error reflects a sampling deficit, we retrain the network from scratch on controlled training sets, each consisting of a fixed base of the 2800 non-hard positions plus $N$ maps drawn from the hard region, for $N \in \{0, 50, 100, 200, 400\}$ and two random draws per $N$.  Accuracy is measured on a fixed set of 120 hard-region positions that are excluded from every training set.  Fig.~\ref{fig:nn_sample_complexity} shows the resulting learning curve.  With no hard-region data the held-out hard-region MAE is $0.149$.  The network cannot extrapolate the sharp boundary structure from the smooth maps it was trained on.  Adding only 50 hard-region maps reduces the error by $61\%$ (to $0.058$), 100 maps by $75\%$, and 400 maps by $89\%$ (to $0.017$).  Each retraining uses an abbreviated fixed-length schedule, so the absolute MAE values are conservative relative to the fully trained production model, and only the trend with $N$ is interpreted.  The error decreases monotonically with $N$, with diminishing returns beyond ${\sim}100$ maps, so the elevated hard-region error reflects a sampling deficit.

\subsection{Transport Integral Validation}

\begin{figure}[!htbp]
  \centering
  \includegraphics[width=\linewidth]{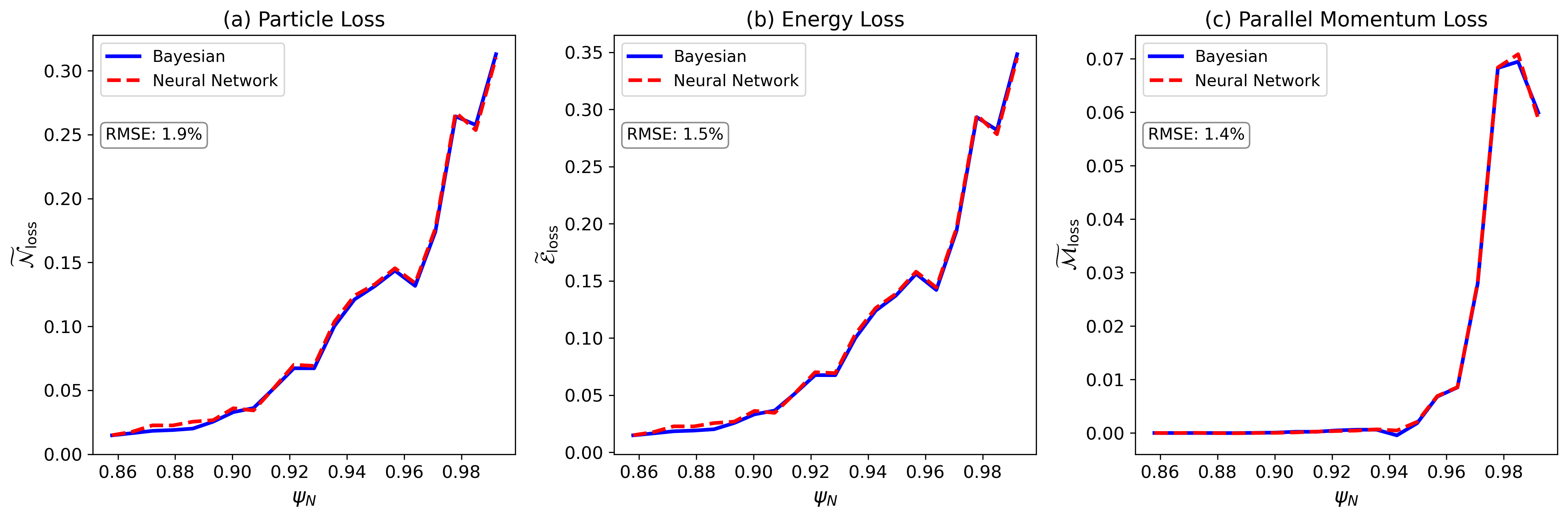}
  \caption{Comparison of IOL transport integrals computed from Bayesian probability maps (solid blue) and neural network predictions (dashed red) as functions of normalized poloidal flux $\psi_N$.  Each integral is evaluated using Eqs.~\eqref{eq:Nloss}--\eqref{eq:Mloss} with ion temperature $kT_i = 1$\,keV, and averaged over the held-out validation and test positions at each radial location.  (a)~Dimensionless lost-ion fraction $\widetilde{\mathcal{N}}_{\mathrm{loss}}$.  (b)~Normalized energy carried by lost ions $\widetilde{\mathcal{E}}_{\mathrm{loss}}$.  (c)~Normalized parallel momentum carried by lost ions $\widetilde{\mathcal{M}}_{\mathrm{loss}}$, capturing the co/counter-current asymmetry relevant to intrinsic rotation.  The RMSE normalized by the maximum Bayesian value, $\mathrm{RMSE}/\max(\widetilde{\mathcal{I}}_{\mathrm{Bayes}})$, is indicated in each panel.}
  \label{fig:nn_integral_comparison}
\end{figure}

The neural network probability maps can be integrated against a Maxwellian distribution using the same quadrature formulas developed in Section~\ref{sec:integrals}.  Fig.~\ref{fig:nn_integral_comparison} compares the particle, energy, and parallel-momentum loss integrals computed from the Bayesian probability maps to those computed from the neural network predictions.  At each radial location $\psi_N$, both the Bayesian and neural network integrals are averaged over the held-out poloidal positions to produce a radial profile.  The comparison is restricted to the 400 held-out validation and test positions, which are never used to update the network weights.  The validation positions set the early-stopping epoch, and the 100 test positions are excluded from training and model selection alike.
The network-derived integrals closely track the Bayesian values across the radial domain sampled by the held-out positions, with RMSE (normalized by the maximum Bayesian value) below $2\%$ for all three moments ($1.9\%$, $1.5\%$, and $1.4\%$ for the lost-ion fraction, energy, and parallel momentum, respectively).  The residual pointwise error concentrated along sharp loss-cone boundaries (Fig.~\ref{fig:nn_hard_region_check}) averages out under the Maxwellian quadrature.  The parallel-momentum integral, which changes sign across the database and is sensitive to the co/counter-current balance, is reproduced at the same level.

\begin{table}[t]
\centering
\caption{Neural-network inference cost for generating $64\times64$ probability maps at different batch sizes.  Timings are measured on a single NVIDIA RTX A5000 GPU for the trained in-domain surrogate.}\label{tab:nn_timing}
\begin{tabular}{lcc}
\toprule
Batch size & NN (GPU) & Per map \\
\midrule
100        & 45\,ms   & 0.45\,ms \\
1000       & 417\,ms  & 0.42\,ms \\
10000      & 4.3\,s   & 0.43\,ms \\
\bottomrule
\end{tabular}
\end{table}
\FloatBarrier

The database was constructed by advancing the Bayesian models at many launch positions together, pooling the trajectories selected at each refinement iteration into one GPU batch.  For one launch position, a standalone run completed the seed stage and four adaptive refinement iterations with the database settings of Table~\ref{tab:hyperparams}, taking about $400\,\mathrm{s}$ to produce one Bayesian probability model.  Table~\ref{tab:nn_timing} reports an average inference time of about $0.4\,\mathrm{ms}$ per probability map for the trained neural network on an NVIDIA RTX A5000 GPU across the tested batch sizes.  Each per-map value is the total runtime divided by the batch size.  These measured workflow costs differ by approximately six orders of magnitude.

\section{Conclusions}\label{sec:conclusions}

This work casts loss-cone boundary determination as an active learning problem in which a Bayesian logistic regression model with RBF features acts as the learner and GC orbit following provides the labels.  For each launch position, trajectories are labeled by whether they hit the wall by $t_\text{max}$, and the resulting labels in $(\KE,\eta)$ space are used to fit the model.  Embedding this model in a quadtree AMR loop concentrates new orbit calculations near the loss-cone boundary and produces a continuous estimate of $P(\text{lost}\mid\KE,\eta)$.  Applied across the sampled edge-region launch mesh, this procedure yields a spatial database of probabilistic loss-cone models for an analytic equilibrium, here exemplified with a DIII-D-like lower single-null solution.

Because each model is a continuous probability map, the database can be integrated against any ion distribution function without further orbit calculations.  Integrating the MAP loss probability against a Maxwellian gives the lost-ion fraction and the normalized energy and parallel momentum carried by lost ions.  The Laplace covariance of each fitted model propagates approximate posterior uncertainty to these moments, again without additional GC simulations.  Across the present edge mesh, the losses concentrate near the separatrix and the outboard side, and the parallel-momentum moment retains a clear co/counter-current asymmetry.  As an example of how the loss-probability database can feed a transport calculation, the orbit-loss torque is balanced against radial momentum diffusion across the flux surfaces of the launch mesh under prescribed transport and renewal coefficients.  The steady rotation is co-current with the plasma current, and the torque is concentrated on the outermost surfaces.  The rotation amplitude depends on the prescribed transport and renewal coefficients.

The neural network surrogate learns the mapping from launch coordinates to loss-probability images and is evaluated orders of magnitude faster than the Bayesian workflow.  On held-out positions, transport moments computed from its predictions closely reproduce the Bayesian values.  The error is largest where a loss-cone map differs strongly from the maps of neighboring launch positions.  A neighbor-difference difficulty score computed from the Bayesian maps locates this hard region before training, and a modest number of additional hard-region maps largely removes the error there.  The score could therefore direct where the Bayesian workflow generates new maps, with multi-fidelity surrogates offering a route to improve accuracy near sharp loss-cone boundaries \cite{zhang2023multi, tatsuoka2025multifidelity} and generative surrogates offering calibrated prediction intervals \cite{sohn2015learning, yang2025conditional, yang2025generative}.

Several limitations and directions for future work remain.  The electric field is zero throughout, so the database omits the modification of the low-energy loss-cone boundary by the radial electric field \cite{chankin1993,miyamoto1996,brzozowski2019}.  The Bayesian fits at different launch positions are independent, so uncertainty bands for spatially averaged profiles are not reported, and dependence between errors at different launch positions has not been modeled.  The per-position cost of the Bayesian active-learning search motivated the neural network surrogate.  A natural extension would run the active-learning search jointly over the four-dimensional domain $(R,Z,\KE,\eta)$, so that the loss-cone boundary is located directly as a surface in that space, spanning launch position and velocity space together.

\section*{Acknowledgments}

This manuscript has been authored by UT-Battelle, LLC, under contract DE-AC05-00OR22725 with the US Department of Energy (DOE). The US government retains and the publisher, by accepting the work for publication, acknowledges that the US government retains a non-exclusive, paid-up, irrevocable, world-wide license to publish or reproduce the submitted manuscript version of this work, or allow others to do so, for US government purposes. DOE will provide public access to these results of federally sponsored research in accordance with the DOE Public Access Plan (http://energy.gov/downloads/doe-public-access-plan).

\bibliographystyle{unsrtnat}
\bibliography{references}

\end{document}